\documentclass[aps,pre,twocolumn,showpacs,superscriptaddress]{revtex4-1}
\usepackage{amsmath,amssymb}
\usepackage{graphics,graphicx,color}
\usepackage{dcolumn,bm}
\usepackage{tikz}
\usepackage{caption}
\usepackage{subcaption}
\usepackage{soul}
\usepackage[colorlinks=true,citecolor=blue,urlcolor=blue]{hyperref}

\def\[{\left[}
\def\]{\right]}
\def\({\left(}
\def\){\right)}
\def\be{\begin{equation}}
\def\ee{\end{equation}}
\def\bea{\begin{eqnarray}}
\def\eea{\end{eqnarray}}

\def\v{\mathbf{v}}

\def\v{{\bm{v}}}

\def\G{{\rm{Im}}\, G}
 \def\d{{\rm d}}

\newcommand{\gaug}
{\affiliation{Institute for Theoretical Physics, Georg-August-Universit\"at G\"ottingen, 37077 G\"ottingen, Germany}}

\begin{document}
\title{Localization properties of the sparse Barrat--M\'ezard trap model}

\author{Diego Tapias}
\email{diego.tapias@theorie.physik.uni-goettingen.de}
\gaug

\author{Peter Sollich}%
\email{peter.sollich@uni-goettingen.de}
\gaug
\affiliation{Department of Mathematics, King's College London, London WC2R 2LS, UK}

\begin{abstract}


  Inspired by works on the Anderson model on sparse graphs, we devise a method to analyze the localization properties of sparse systems that may be solved using cavity theory. We apply this method to study the properties of the eigenvectors of the master operator of the sparse Barrat--M\'ezard trap model, with an emphasis on the extended phase. As probes for localization, we consider the inverse participation ratio and the correlation volume, both dependent on the distribution of the diagonal elements of the resolvent. Our results reveal a rich and non-trivial behavior of the estimators across the spectrum of relaxation rates and an interplay between entropic and activation mechanisms of relaxation that give rise to localized modes embedded in the bulk of extended states. We characterize this route to localization and find it to be distinct from the paradigmatic Anderson model or standard random matrix systems.
\end{abstract}

\maketitle

\section{Introduction}

Trap models were introduced in the literature to provide a coarse--grained description of the dynamics in glassy systems~\cite{bouchaud1992weak, dyre1995energy}. These models exhibit the essential properties of glassy materials, which are:  a glass transition temperature below which the system falls out of equilibrium, a slow (compared to an exponential) decay of the energy with time, and \emph{aging}~\cite{arceri2020glasses}. Currently, there is a wide spectrum of models that can be considered trap--like, ranging from mean--field to sparse (in terms of connectivity), from discrete to continuous (in terms of the configuration space) and with additional features such as correlation between trap depths and node degrees or the inclusion of dynamical facilitation~\cite{barrat1995phase, moretti2011complex, margiotta2018spectral, carbone2020effective, scalliet2021excess}. 

The ingredients needed to define a trap model are: a network of configurations,  an energetic landscape that assigns energies to these configurations, and a set of transition rates that respect detailed balance with respect to the Boltzmann distribution. Historically, models with networks of mean field type, i.e.\ fully connected, were the first to be analyzed~\cite{dyre1987master, bouchaud1992weak, barrat1995phase,dyre1995energy}. 
Remarkably, these simplified cartoons of the energy landscape and its topology were rich enough to capture some of the essential features of real glasses. Regarding the choice of microscopic transition rates, two model variants are most widely studied: the Bouchaud trap model~\cite{bouchaud1992weak} (sometimes simply called trap model) and the Barrat--M\'ezard (BM) trap model~\cite{barrat1995phase} (also known as step model). In the first, the transition rates are given by an Arrhenius formula, which models activation across energetic barriers located at some threshold level. For the second, the transition rates are of Glauber form and so encode the effects on the motion of both entropic and energetic barriers. The dynamical properties of those models have been analyzed in depth both from a formal mathematical point of view and by extensive numerical simulations~\cite{ bertin2003cross, bovier2005spectral, arous2008arcsine, cammarota2015spontaneous}.

In recent years there has been {additional} interest in trap models because they have been shown to be relevant for the description of the dynamics of classical disordered systems such as the REM (random energy model) or the Ising $p$--spin model~\cite{baity2018activated, baity2018activated1, stariolo2019activated, stariolo2020barriers}. Moreover, an extension of the Bouchaud trap model that includes interactions between traps in the context of kinetic facilitation has been successful in explaining the excess wings observed in the dynamical susceptibility spectra of real glasses~\cite{scalliet2021excess}.

In a series of recent studies~\cite{margiotta2018spectral, margiotta2019glassy, tapias2020entropic}, the role of sparsity of the networks for the Bouchaud and BM models has been investigated. For those sparse networks the trap model is formulated as a continuous time Markov chain with a master operator that depends on the connectivity of the network and the landscape. An analytical approach that is suitable for studying such models is the cavity method~\cite{rogers2008cavity, mezard2009information, susca2021cavity}, which provides access to both spectral and time--domain properties. The overall outcome is that sparsity yields non-trivial features in the dynamics that translate into richer and more realistic physics. In particular, the sparse BM model exhibits a crossover from an initial transient entropic--driven relaxation to a long time activated dynamics as a consequence of the finite connectivity of the network and the eventual lack of ``escape directions''  towards lower energies~\cite{tapias2020entropic}. This picture is very intuitive if one thinks of the relaxation in a system of particles with short--range interactions from some suitably random initial condition: 
rearrangements first take place as a downhill motion in the energy landscape driven by entropic features, while further relaxation implies crossing of energy barriers activated by thermal fluctuations. The crossover between these two types of dynamics may be relevant for describing relaxation in models that  present qualitatively similar features, for instance the Random Orthogonal Model~\cite{crisanti2000activated}.

In this work, we {devise a method to  study the localization properties of the eigenmodes of the (symmetrized) master operator, inspired  by} techniques used for the analysis of the Anderson Model on Random Regular Graphs~\cite{biroli2010anderson, biroli2012difference, biroli2018delocalization,tikhonov2019statistics, tikhonov2019critical, tikhonov2021anderson}. To do so we rely on the population dynamics algorithm~\cite{kuhn2008spectra, tapias2020entropic}. We show that the infinite size self--consistent equations for the cavity Green's functions may be used to understand the statistics of eigenfunctions belonging to the extended regime.  Essentially, we construct a simple {scheme} to mimic the mean behavior of finite size network instances that allows us to predict the mean Inverse Participation Ratio (IPR) of the extended eigenvectors, and to estimate the correlation volume associated with a given relaxation mode.

Our method is appealing because it {works} beyond the limit in which direct diagonalization of the master operator is possible, allowing us to access a regime where even the more efficient single instance cavity method would demand very heavy computational resources. {One of our main results is the characterization of the extended regime in terms of a well--defined correlation volume, extracted from the thermodynamic limit but crucial in the behavior of finite instances. Moreover, we validate an estimator of the IPR that may be used to predict numerical IPR values of finite size instances in the extended regime (and not only their scaling) via the cavity method.  Beyond these methodological advances, we demonstrate the existence of localized states within the bulk of extended states for finite temperatures; these states appear due to the entropic dynamics that characterizes the low temperature phase of the BM model. The mechanism of localization that generates them is different from others found in the context of Random Matrix Theory and is in itself an interesting result. Finally, we show that the localization probes across the spectrum of relaxation rates are non-monotonic and, in general, more complex than for the sparse Bouchaud trap model.} 

The paper is organized as follows. In section~\ref{sparsebm} we introduce the sparse Barrat--M\'ezard trap model and discuss its main spectral properties. {In section~\ref{estimators} we introduce the relevant estimators for the analysis of localization properties in sparse systems.  In section~\ref{method} we present our method.}  Section~\ref{results} has our main numerical results. Finally, in section~\ref{concs} we summarize our results and the key features of our method, and set out some open research questions to be addressed in the future. 

\section{Sparse Barrat--M\'ezard trap model} 
\label{sparsebm}

The sparse BM trap model is a Markov chain with a  configuration space that is a sparse network, transition rates of Glauber form and random trap depths (node energies) that reflect the rough nature of the glassy landscape. For the sparse network we consider specifically a random regular graph (RRG), where every node has exactly $c$ neighbours. The model can be defined via the master operator $\bf{M}$ with elements
\begin{align}
  M_{ji} = \frac{A_{ji}}{c} \frac{1}{1 + \exp(\beta(E_j - E_i))} \, , \quad M_{ii} = - \sum_{j \neq i} M_{ji}
  \label{mast}
\end{align}
Here $M_{ji}$ is the rate for transitions from node {(configuration)}  $i$ to node $j$ and $-M_{ii}$ is the escape rate from node $i$. $\bf{A}$ denotes the adjacency matrix of the graph,  with $A_{ij}=A_{ji}=1$ indicating the presence of an edge and $A_{ij}=0$ otherwise, $c$ is the connectivity of the network (in general given by $c = \frac{1}{N} \sum_{i,j} A_{ij}$), and $E_i$ is the energy depth of node $i$. We exclude self-edges, setting $A_{ii}=0$. The trap depths $E_i$ are sampled independently from an exponential distribution $\rho_E(E) = \frac{1}{T_{\rm{g}}} {\rm{e}}^{-E/{T_{\rm{g}}}}$.  The dimensions of the matrices are $N \times N$ where $N$ is the size of the network. Finally, $\beta=1/T$ is the inverse temperature as usual.

The stationary solution of the master equation $\dfrac{d {\bm{p}}(t)}{d t} = {\bf{M}} {\bm{p}}(t)$ is a probability vector with the Boltzmann weights as entries, i.e. ${\bm{p}}^{\rm{eq}} \propto ({\rm{e}}^{\beta E_1}, \ldots, {\rm{e}}^{\beta E_N} )$. The fingerprint of trap models is that in the limit $N \to \infty$ this distribution becomes non-normalizable for $T < T_{\rm{g}}$, which identifies $T = T_{\rm{g}}$ as the glass transition temperature. From now on we will work in units where $ T_{\rm{g}} = 1$.

The stationary solution allows us to symmetrize  ${\bf{M}}$ as follows: ${\bf{M}}^s = {\bf{P}}_{\rm{eq}}^{-1/2} {\bf{M}}  {\bf{P}}_{\rm{eq}}^{1/2} $ where $ {\bf{P}}_{\rm{eq}}$ is a diagonal matrix with elements $( {\bf{P}}_{\rm{eq}})_{ii} \propto {\rm{e}}^{\beta E_i}$. This symmetrization, which relies on the fact that the transition rates obey detailed balance, does not change the eigenvalues of $\bf{M}$; it also does not qualitatively change the relevant features of the eigenvectors (see also the discussion for the sparse Bouchaud trap model in Ref.~\cite{margiotta2018spectral}). From now on, whenever we refer to the master operator we mean ${\bf{M}}^s$ unless explicitly stated otherwise. 

Our primary interest lies in the eigenvectors of the master operator, which we will refer to as states. Physically, these states correspond to the relaxation modes of the system. Their {associated} eigenvalues $\lambda$ are negative and give the relaxation rates $-\lambda$. Knowledge of the distribution of eigenvalues (density of states) and the statistics of the eigenvectors provides essentially all the information about the dynamics of the system. To access these spectral properties we employ the cavity method (see details in Ref.~\cite{tapias2020entropic}), which for a given instance of the disorder -- in trap depths and network structure -- is based on the solution of the following system of equations
\begin{align}
\omega_k^{(j)} = i \lambda_\epsilon  {\rm{e}}^{\beta E_k}c  + \sum_{l \in \partial k \setminus j} \frac{i K(E_k, E_l) \omega_{l}^{(k)} }{i K(E_k, E_l) + \omega_l^{(k)} } \, ,
\label{cavities}
\end{align}
{with $\lambda_\epsilon \equiv \lambda - i \epsilon$},  $\epsilon>0$ a small regularizer  and
 \begin{align}
  K(E_i, E_j) = \frac{{\rm{e}}^{\beta(E_i + E_j)/2}}{2\cosh(\beta(E_i - E_j)/2)}
\end{align}
The equations~\eqref{cavities} are the Single Instance (SI) cavity equations, and $\omega_k^{(j)}$ is the ``cavity precision'' for node $k$ in the absence of node $j$, i.e.\ for a modified network with a cavity cut out around $j$. After solving eq.~\eqref{cavities}, one inserts the solution for the cavity precisions into the expression for the ``marginal precisions'', which is
\begin{align}
\omega_j = i \lambda_\epsilon {\rm{e}}^{\beta E_j}c +  \sum_{k \in \partial j} \frac{i K(E_j, E_k) \omega_k^{(j)} }{i K(E_j, E_k)  + \omega_k^{(j)}  } 
\label{marginals}
\end{align}
It is common in the disordered systems literature to express results in terms of the Green's functions or marginal variances $G_{jj}$, which are the diagonal entries of the resolvent matrix
\begin{align}
    {\bf{G}}(\lambda_\epsilon) = \left( \lambda_\epsilon {\bf{I}}  - {\bf{M}}^s \right)^{-1} \, ,
    \label{resmatrix}
\end{align}
with $\bf{I}$ the identity matrix. With the scaling used in eqns.~\eqref{cavities} and~\eqref{marginals}, the relation between the $G_{jj}$ and the marginal precisions is
\begin{align}
  G_{jj} = \frac{i {\rm{e}}^{\beta E_j} c}{\omega_j}
\end{align}
Once the marginal variances are known for every node of the network, one may estimate the density of states for the master operator $\bf{M}^s$ via the identity~\cite{susca2021cavity, livan2018introduction}  
\begin{align}
  \rho(\lambda) = \lim_{\epsilon \to 0} \frac{1}{\pi N} \sum_{i=1}^N \G_{ii}
 \label{sidos}
\end{align}
where ${\rm{Im}}$ indicates the imaginary part. From the expression~\eqref{sidos} one can also read the DOS as the average of the local Density of States (local DOS) defined at node $i$ as
\begin{align}
  \rho_i = \lim_{\epsilon \to 0} \frac{1}{\pi} \G_{ii}
  \label{ldos}
\end{align}
In the large $N$ limit the variation among results for individual single network instances is expected to go to zero because of self--averaging. For $N\to\infty$ one can then exploit the structure of the cavity equations~\eqref{cavities} to arrive at a self--consistent equation for the joint distribution of local energies and cavity precisions $\zeta(\omega, E)$: 
\begin{align}
    { \zeta(\omega, E) = \rho_E(E) \int \prod_{ j=1}^{c-1} \d E_j \d \omega_j \zeta(\omega_j, E_j) \delta( \omega - \Omega_{c-1}) }
 \label{selfc}
\end{align}
with 
\begin{align}
{ \Omega_{c-1}(\{\omega_l, E_l\}, E) = i\lambda_\epsilon {\rm{e}}^{\beta E} c +  \sum_{l=1}^{c-1} \frac{i K(E, E_l) \omega_{l}}{i K(E, E_l) + \omega_l} }
\label{omegas}
\end{align}
 Equation~\eqref{selfc} is solved with a Population Dynamics (PD) algorithm {(also known as ``pool method'')} that iteratively updates members $(\omega,E)$ of a population of a certain size representing $\zeta(\omega,E)$~\cite{kuhn2008spectra, monthus2008anderson, tapias2020entropic}. The local Green's function then becomes the random variable
\begin{align}
   G = \frac{i {\rm{e}}^{\beta E} c}{\Omega_c}
\end{align}
Here $\Omega_c\equiv \Omega_c(\{\omega_l,E_l\},E)$ is a function of $c$ independent samples from $\zeta(\omega_l,E_l)$, and $E$ drawn from $\rho_E(E)$. This can be used to estimate the distribution $P(G)$, and more specifically $P(\G)$, for a given value of $\epsilon$. The DOS (eq.~\eqref{sidos}) in this setting becomes
 \begin{align}
   \rho(\lambda) = \lim_{\epsilon \to 0} \frac{1}{\pi} \langle \G \rangle
   \label{DOS}
  \end{align}
For a detailed analysis of the DOS for the BM model we refer the reader to Ref.~\cite{tapias2020entropic}, where a systematic analysis of the spectral density in the $(T, c)$-parameter space is presented. We only recap some relevant features here:
 \begin{itemize}
 
\item For finite $c$, $T < 1$ and $\lambda  \to 0$, $\rho(\lambda) \sim |c\lambda|^{T-1}$.
 
\item For finite $c$, and $T < 1/2$, the DOS diverges as a power law at $\lambda= -1/c$ as $\rho(\lambda) \sim |\lambda+1/c|^{2T-1}$. In general,  at the values $\lambda_k^* = -k/c$ with $k = \{0, 1, \ldots, c\}$, the DOS diverges as a power law with a temperature dependent exponent.
 
\item In the zero temperature limit $T = 0$, the relaxation rates become the escape rates  and the spectrum is a superposition of Dirac $\delta$-functions at the values $\lambda_k^*$. The relaxation modes are {completely} localized on individual local configurations of the landscape {(see Appendix~\ref{locsec})}.

  \item In the mean--field limit, $c \to \infty$, the relaxation rates also become identical to the escape rates, leading to a spectrum that is flat for $\lambda \to 0$, with a limiting value $\rho(\lambda\to 0) = \pi T/\sin(\pi T)$.
 
\item In the infinite temperature limit, $T \to \infty$, the DOS follows the Kesten--McKay law~\cite{kesten1959symmetric, mckay1981expected, bauerschmidt2019local}
  and all the states 
  {are extended (or delocalized) with high probability (see Ref.~\cite{dumitriu2012sparse} and references therein)}. In this regime, the master operator (eq.~\eqref{mast}) becomes the scaled Laplacian matrix of the associated graph {and the master  equation describes an (unbiased) random walk in the network. Moreover, the associated eigenvectors have Gaussian distributed entries~\cite{clark2018moments, backhausz2019almost}.} 

\end{itemize}
 
Before leaving this section, we comment on the effects of $\epsilon$ on the DOS and their relation with localization. Usually one chooses a small value of $\epsilon$, say of order $\sim 10^{-4}$ to compute~\eqref{DOS} and the result is reported as the (total) density of states. However, as pointed out already in the classical reference~\cite{abou1973selfconsistent}, in the limit $\epsilon \to 0$, $P(\G)$ for a given $\lambda$ can be either regular, i.e. non-trivial, or singular, i.e. $\delta(\G)$. In the first case, the corresponding {state} is extended, while in the second case it is localized. Thus, in principle, one could distinguish the nature of the states by computing the (extended) DOS for a given population in the limit $\epsilon \to 0$. This was done (with $\epsilon = 10^{-300}$) for the sparse Bouchaud trap model in Ref.~\cite{margiotta2018spectral} to identify the \emph{mobility edges}, i.e.\ the points near the edges of the spectrum where a transition from extended to localized states takes place. A drawback of this method, however, is a strong population size dependence.

\section{Estimators}
\label{estimators}

We aim to study the localization properties of the eigenvectors of the master operator. Intuitively, a state is considered localized if its entries are concentrated on few nodes in the large $N$ limit. On the other hand, the state is extended if its entries are roughly equally distributed in the same limit. A standard probe to distinguish between those scenarios is via the scaling of the Inverse Participation Ratio (IPR). This quantity, computed as
\begin{align}
  I_2(\bm{v}_\alpha) = \sum_{i=1}^N \bm{v}_{\alpha,i}^4
  \label{empiripr}
\end{align}
for the vector $\bm{v}_\alpha$ is a measure of how (un-)evenly the squared norm of a vector is distributed across its entries~\cite{livan2018introduction}.

In general, the IPR scales as $N^{-\mu}$,  with $0 \leq \mu \leq 1$. The extreme cases $\mu = 0$ and $\mu = 1$ correspond to localized and extended (or delocalized) vectors, respectively.  It is important to remark that the scaling should be asymptotic in $N$ and one may reach incorrect conclusions if one focuses on small graphs (as highlighted in e.g.~\cite{tikhonov2016anderson}).  {In the extended regime, the inverse of the prefactor $\gamma$ in $I_2 \approx \gamma/N$ quantifies how extended the states are. A classical ensemble of delocalized modes, the Gaussian Orthogonal Ensemble~\cite{potters2020first}, yields states with $\gamma =3$, which is the analytical result for vectors with Gaussian distributed entries. This is also the situation for the $T \to \infty$ limit of our model, as reviewed in the previous section. We will refer to such states as ``fully extended''. }

From the point of view of the cavity method, with which one has access to information on the resolvent entries rather than on individual eigenvectors, one estimates the mean IPR for a given $\lambda$ in the spectrum~\cite{mirlin2000statistics, metz2010localization, biroli2012difference, tikhonov2019statistics, tikhonov2019critical}. In this paper we will use the estimator
\begin{align}
  I_2(\lambda) =  \frac{3}{N}  \frac{\langle (\G)^2 \rangle }{\langle \G \rangle^2}
       \label{iprtikh}
\end{align}
that is suitable for extended states~\cite{tikhonov2019critical}. {In equation~\eqref{iprtikh} and subsequent expressions of this section we implicitly consider the small $\epsilon$ limit.}   We note that the expression~\eqref{iprtikh} is not the most common estimator for the IPR in the context of cavity theory. In appendix~\ref{iprest} we compare it with the standard one and show why eq.~\eqref{iprtikh} performs better in the extended regime.

Beyond the IPR, an important probe of the nature of the eigenmodes is the correlation volume $N_\xi$. This quantity is defined only in the extended regime and characterizes the size of the neighborhood around a typical node within which an eigenvector will appear to be localized,
so that its squared norm (or the local DOS) within that neighborhood is concentrated on a few adjacent nodes. In the high $T$ regime where the vectors are fully extended, the correlation volume is $O(1)$. 
As the temperature decreases, $N_\xi$ increases and then diverges at the localization transition. This quantity is defined up to an unknown prefactor by the inverse of the typical value of the DOS~\cite{tikhonov2019critical, tikhonov2021anderson}, ${\rm{e}}^{\langle \ln \G \rangle}$. We set the prefactor to ensure the required scale invariance~\footnote{In Ref.~\cite{tikhonov2019critical}, $N_\xi$ is introduced as the inverse of the typical value of the DOS. However, this violates the requirement that the expression for $N_\xi$ should be invariant under rescaling of ${\bf{M}}^s$, $\lambda$ and $\epsilon$ by the same factor.}, giving
\begin{align}
  \label{cvol}
  N_{\xi} = {\langle {\rm{Im}} G \rangle }{{\rm{e}}^{-\langle \ln {\rm{Im}} G \rangle}}
\end{align}

A way to make sense of the correlation volume is by examining a spatial two--point correlation function. First, we notice that on a RRG a neighborhood of radius $r$ has a volume scaling as $r^{c-1}$ where $c$ is the connectivity. Inverting this, we can introduce the correlation length
\begin{align}
  r_\xi = \log_{(c-1)}{(N_\xi)}=\frac{\ln N_\xi}{\ln(c-1)}
  \label{corrlength}
\end{align}
We then expect (and confirm numerically in section~\ref{results}) that this length controls the decay of the correlation of the local DOS~\cite{efetov1992scaling}, defined as
\begin{align}
  \kappa(r) = \frac{\left \langle \rho_i \rho_j  \right \rangle}{\rho^2 } = \frac{\langle {\rm{Im}} G_{ii}\, {\rm{Im}} G_{jj} \rangle}{\langle {\rm{Im}} G \rangle^2}
  \label{corr1}
\end{align}
and its connected version
\begin{align}
  \kappa^c(r) =  \frac{\langle {\rm{Im}} G_{ii}\, {\rm{Im}} G_{jj} \rangle}{\langle {\rm{Im}} G \rangle^2} - 1
  \label{corr}
\end{align}
The averages here are over all pairs of nodes $(i,j)$ that have distance $r$.
Notice that at $r = 0$, $\kappa^c(0) = N I_2/3 -1 $ (cf.~eq.~\eqref{iprtikh}). For a similar set of correlation functions we refer the reader to Ref.~\cite{tikhonov2019statistics}.

\section{Method}
\label{method}

We will rely on population dynamics to estimate the distribution $P(\G)$ generated by instances of size $N$ in the extended regime, and use this distribution to evaluate the estimators of the previous section. More specifically, we numerically solve equation~\eqref{selfc} for a value of $\epsilon$ that is determined by the size $N$ of the instances we are interested in studying as
\begin{align}
  \epsilon = \frac{C}{\pi \rho(\lambda) N} 
  \label{gene}
\end{align}
with $C/\pi$ an $O(1)$ constant, {and then evaluate the estimators~\eqref{iprtikh} and~\eqref{cvol}}. In this way, we can access sizes bigger than those obtained using the most efficient numerical routines to solve the SI cavity equations~\eqref{cavities}. {Here we stress that within the PD algorithm, the population size $N_p$ is in general different from (and usually smaller than) $N$; thus, within PD, $N$ should be thought of as an effective instance size.
}

{The justification for the scaling~\eqref{gene} and the use of population dynamics comes from the following observations for the Anderson model, on the same type of networks as considered here, in its extended phase~\cite{abou1973selfconsistent, biroli2018delocalization, kravtsov2018nonergodic, tikhonov2019statistics, tikhonov2019critical}: 
  (i) The mean of the product $NI_2$ (as obtained from Direct Diagonalization (DD), see below) becomes $N$ independent for sizes much larger than the correlation volume and can be estimated with PD via equation~\eqref{iprtikh}. (ii) For a given $\epsilon$, both PD and SI with sufficiently large $N$ provide estimates for $P(\G)$, with PD generally giving more accurate results.
  (iii) For a given finite instance, the results for the typical and mean value of the distribution $P(\G)$, from Exact Diagonalization (ED) and from the SI cavity method agree whenever the regularizer $\epsilon$ is of the order of the mean level spacing, i.e.\ $\epsilon \sim 1/(\rho N)$ (see also appendix~\ref{agreement}).} For future reference, by ED we mean the numerical evaluation of the resolvent (eq.~\eqref{resmatrix}) using its expansion in terms of the eigenvectors and eigenvalues of the symmetrized master operator (see eq.~\eqref{resmatrixexp} in {appendix~\ref{agreement}}) while we refer to the  calculation of the IPR for individual eigenvectors 
as Direct Diagonalization (DD). 

{In short, the estimation of $P(\G)$ with PD and the scaling~\eqref{gene} is meaningful because of the convergence of the distribution for large $N$ and small $\epsilon$ whenever the eigenvectors are extended.}

As an additional point, we highlight that even though we use a finite population of size $N_p$ for the PD algorithm, we can estimate the mean properties of finite instances of size $N$ well above $N_p$. {Our limitation is the actual number of independent samples from $P(\G)$ that we can generate for a given population. We restrict our method to $10^3 \times N_p$ samples.} For technical details about how to implement PD for the BM model, we refer the reader to Ref.~\cite{tapias2020entropic}. We only mention in passing that compared to the population size used to estimate the DOS~\cite{tapias2020entropic} we use bigger populations in this work, usually $N_p = 10^5$. 

Finally, two remarks regarding our choice for $\epsilon$ (eq.~\eqref{gene}) are in order. {First,  the exact value of the constant $C$ is not relevant as long as $\epsilon \sim 1/(\rho N)$ because above a certain $N$ the PD (or SI) results would become $\epsilon$ independent. As a matter of fact, $C$ is only relevant for estimating $P(\G)$ with ED as discussed in Appendices~\ref{agreement} and~\ref{iprest}. For definiteness, we will fix  $C = 3$.} Second,  from formula~\eqref{gene} one observes that the DOS or, in other words,  $\langle \G \rangle$ is needed to compute $\epsilon$; yet this average is an {\em output} of the method and not known {\em a priori}. We deal with this by providing an initial estimate for $\rho(\lambda)$ using PD with a small population size ($N_p = 10^3$ or $10^4$) and  a fixed initial value of $\epsilon = 10^{-4}$, and then re-assign $\epsilon$ using~\eqref{gene}. The slight fluctuations in the resulting value of $\epsilon$ that come from the numerical uncertainty in the initial estimate for $\rho(\lambda)$  are immaterial because of the aforementioned irrelevance of the $C$ value for the cavity results.
 
 \section{Results}
\label{results}

In this section, we present support for the correctness of our method and numerical results for the estimators introduced in section~\ref{estimators}. To see the deviations from mean field networks most clearly we fix the connectivity to the lowest meaningful value, namely, $c=3$. This ensures that the network is connected, with the fraction of nodes outside the giant connected component vanishing for large $N$~\cite{bollobas2001random}, and leaves temperature $T$ as our main control parameter.

{We recall that for fixed finite connectivity, the high $T \to \infty$ limit produces fully extended states, with $N I_2 = 3 $ (see Ref.~\cite{clark2018moments} for a systematic study of the moments of the IPR). On the other hand, the limit $T = 0$ gives fully localized states at the eigenvalues $\lambda^*_k = -k/c$ associated with the entropic relaxation mechanism (see appendix~\ref{locsec}). In our analysis, we will focus on the localization properties of the spectrum as we decrease temperature, hence effectively increase the amount of disorder in the system.}

\begin{figure}
   \includegraphics[width=\linewidth]{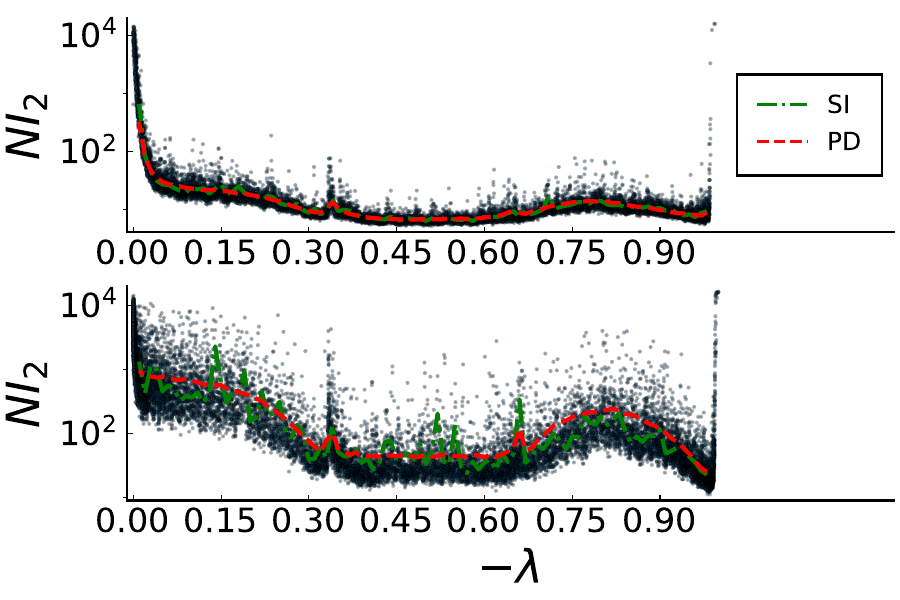} 
  \caption{Inverse Participation Ratio (multiplied by $N$). Scatter plot: Direct Diagonalization (eq.~\eqref{empiripr}).  Comparison with estimator eq.~\eqref{iprtikh}, obtained either by  Single Instance (dash-dotted line) or Population Dynamics (dashed line). Size of the network $N=2^{14}$, population size $N_p = 10^5$ with the same effective $N = 2^{14}$. Temperature $T = 0.9$ (top), $T = 0.5$ (bottom),  eigenvalue grid size $\Delta \lambda = 0.01$.}
   \label{iprsied}
 \end{figure}

\begin{figure}
  \centering
  \begin{subfigure}[t]{0.4\textwidth}
    \caption{}
    \centering
    \includegraphics[width=\linewidth]{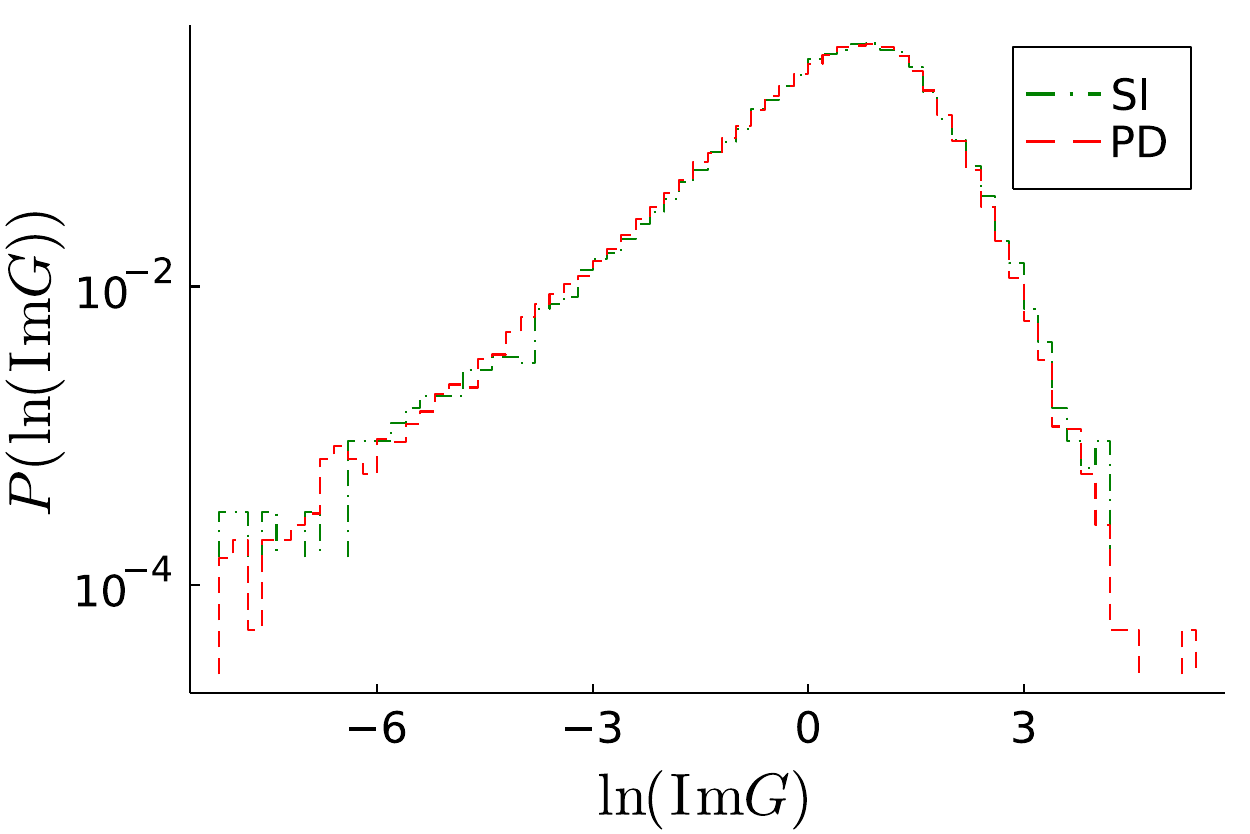} 
  \end{subfigure}
  \hfill
  \begin{subfigure}[t]{0.4\textwidth}
    \caption{}
    \centering 
    \includegraphics[width=\linewidth]{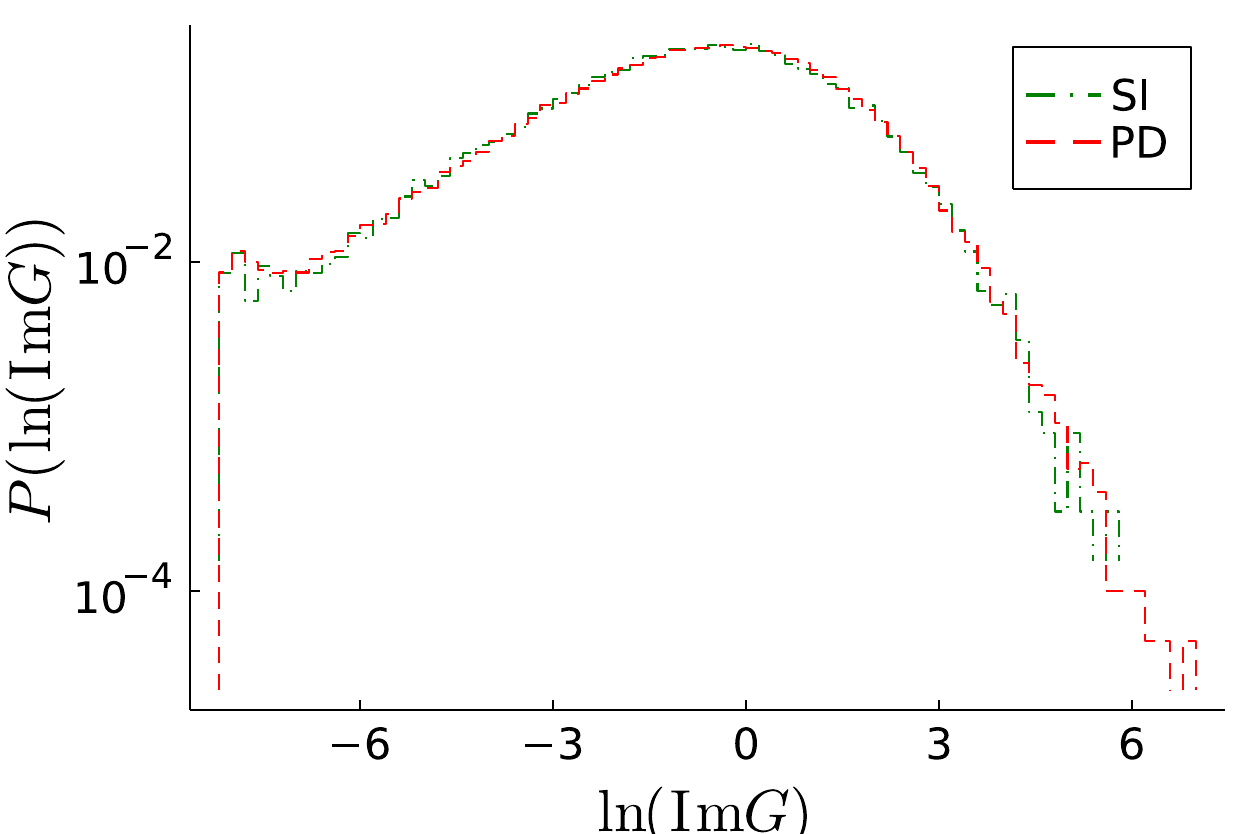} 
  \end{subfigure}
  \caption{Histograms of the full distribution $P(\G)$ (after a logarithmic transformation) for $\lambda = -1/2$. Temperature $T = 0.9$ (top), $T = 0.5$ (bottom). Single Instance (dash-dotted) for a network with size $N = 2^{14}$. Population Dynamics (dashed) with the same effective $N = 2^{14}$ and population size $N_p = 10^3$.}
  \label{sivspd}
\end{figure}

We start by showing the agreement between the mean IPR obtained via eq.~\eqref{iprtikh} using Population Dynamics (PD), Single Instance cavity method (SI) and Direct Diagonalization (DD, eq.~\eqref{empiripr}). In Figure~\ref{iprsied} we compare two temperatures for a system of size $N = 2^{14}$, close to the largest size where we can still use DD. {In the figure} one can appreciate that the estimator given by~\eqref{iprtikh} performs well not just qualitatively but also quantitatively. The figure also supports the idea that one can use the infinite--size self--consistent equation~\eqref{selfc} for predicting the average behavior of instances of size $N$. Since ultimately our estimators are based on $P(\G)$, we show this distribution explicitly in figure~\ref{sivspd} at a fixed value of $\lambda$, comparing estimates from SI and PD for two different temperatures. The agreement between the methods is clear; furthermore it is evident that PD performs better in estimating the tails of the distribution {even though in this case $N_p < N$}. 

\begin{figure}
  \centering
  \begin{subfigure}[t]{0.4\textwidth}
    \caption{}
    \centering
    \includegraphics[width=\linewidth]{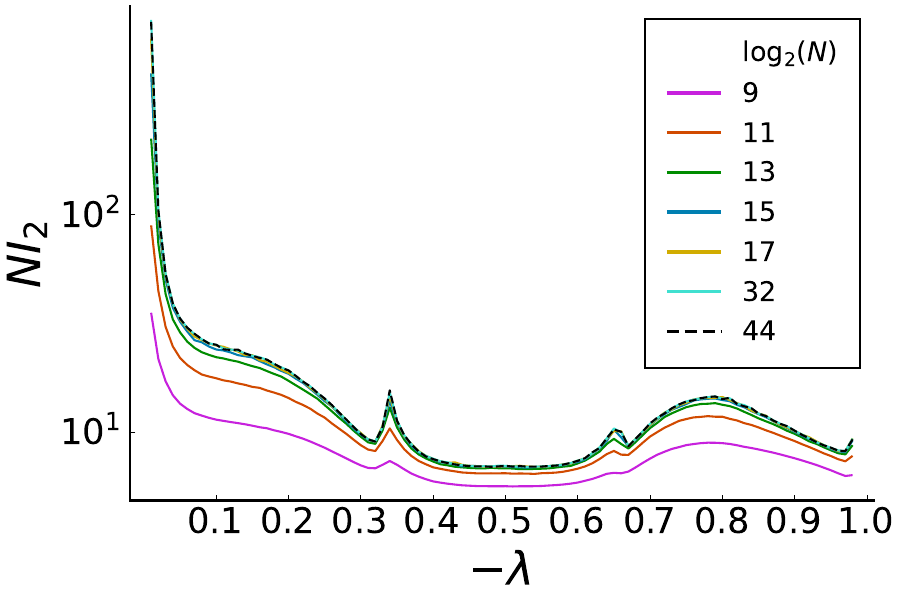} 
  \end{subfigure}
  \hfill
  \begin{subfigure}[t]{0.4\textwidth}
    \caption{}
    \centering
    \includegraphics[width=\linewidth]{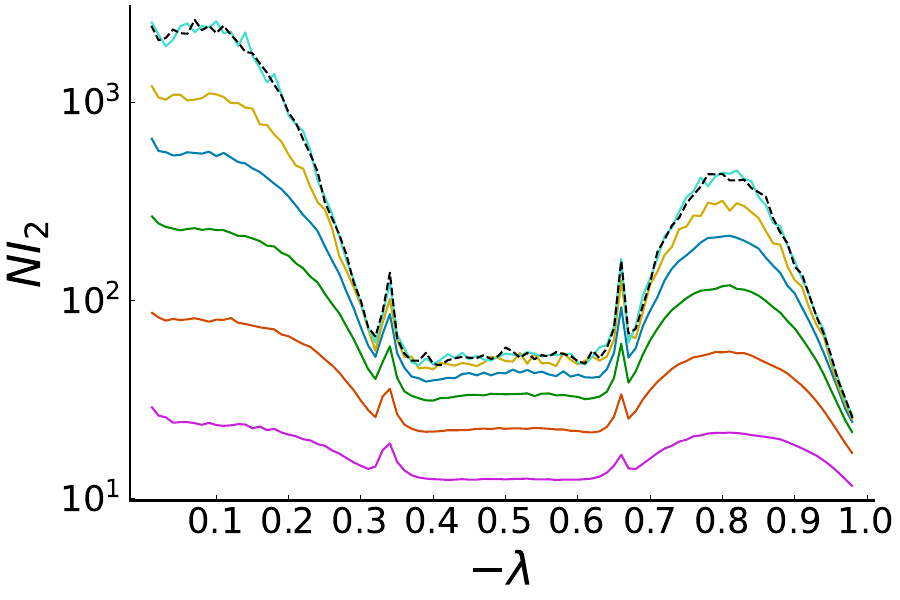} 
  \end{subfigure}
  \caption{$N$-dependence of the product $N I_2$ for two different temperatures. Top: $T = 0.9$, bottom: $T = 0.5$. Eigenvalue grid size $\Delta \lambda = 0.01$. Population size $N_p = 10^4$. Results are averages over one hundred independent runs.}    
  \label{niprscaling}
\end{figure}

In Figure~\ref{niprscaling} we show the product $N I_2$ for a relatively high and low temperature. {From now on we will use $N$ to indicate the effective system size that we are exploring via PD unless otherwise stated.} The first feature that one notices is the non-monotonicity of $NI_2$ across the spectrum for both temperatures. As regards the $N$-dependence, for the higher temperature, we observe that the results saturate at moderate values of $N$ between $2^{13}$ and $2^{15}$ across most of the spectrum of eigenvalues.  Close to the equilibrium state ($\lambda = 0$), however, we do not see saturation. This is consistent with the existence of \emph{mobility edges} at both ends of the spectrum where a localization transition occurs (see for instance, Ref.~\cite{margiotta2018spectral}). We also notice the peaks around the values $-\lambda \in \{ 1/3, 2/3 \}$, separated by a region with the lowest values of $NI_2$. This feature is intriguing and suggests that around the peaks, the relaxation modes have unique properties. We perform a more systematic analysis below. 
{In addition, we notice that for the lower temperature (Fig.~\ref{niprscaling} bottom), there is no saturation up to {$N=2^{17}$, in contrast with the high temperature case,} with further differences in behavior depending on the region of the spectrum. {The figure suggests that {at $N=2^{17}$} the central region is closest to saturation, followed by the region of fast modes (high $|\lambda|$) and finally the region of slow modes.} {{
For much larger values of $N=\{2^{32}, 2^{44}\}$, saturation across the full range of $\lambda$ is then observed.}}
}

\begin{figure}
  \centering 
  \begin{subfigure}[t]{0.4\textwidth}
    \caption{}
    \centering
    \includegraphics[width=\linewidth]{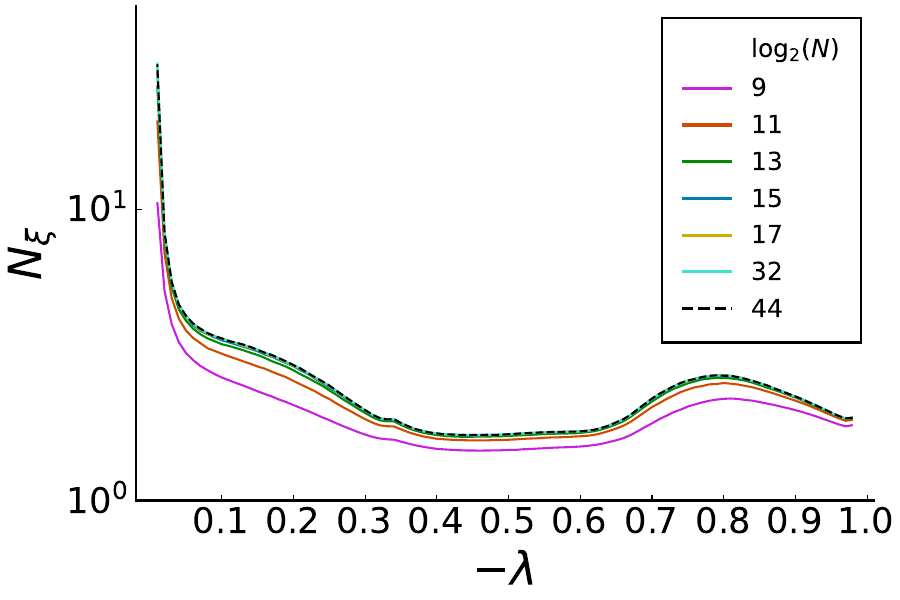}  
  \end{subfigure}
  \hfill
  \begin{subfigure}[t]{0.4\textwidth}
    \caption{}
    \centering
    \includegraphics[width=\linewidth]{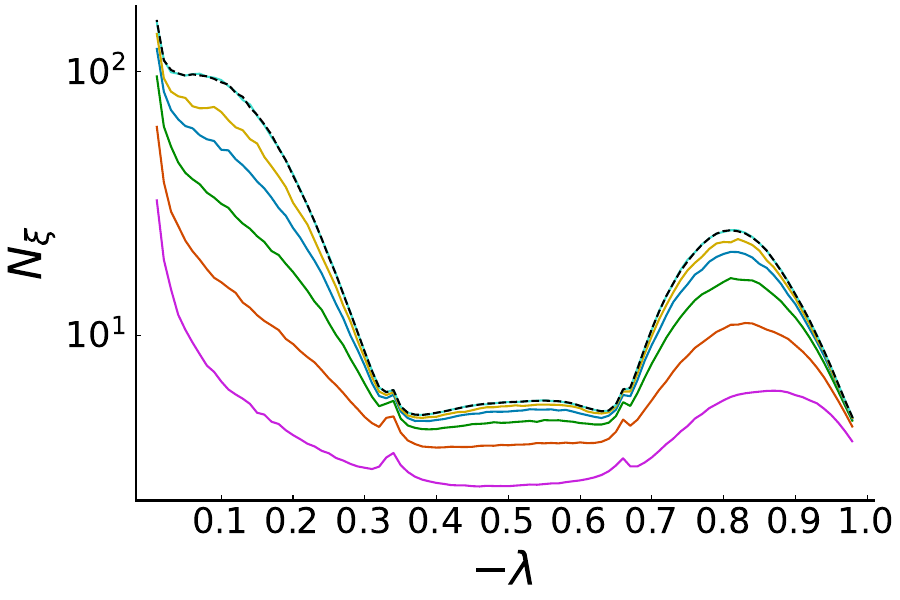} 
  \end{subfigure}
  \caption{$N$-dependence of the correlation volume $N_\xi$ for two different temperatures. Top: $T = 0.9$, bottom: $T = 0.5$.  The raw data are the same as used for Fig.~\ref{niprscaling}. }
  \label{corrscalingp}
\end{figure}

We proceed with the estimation of the correlation volume $N_\xi$ for the same set of temperatures. Figure~\ref{corrscalingp} shows the results, which are derived from the same data used to generate Figure~\ref{niprscaling}. We observe again non-monotonic behaviour across the spectrum; the values of $N_\xi$ are smaller in absolute terms than those of $N I_2$, {and the $\lambda$-dependence is smoother} with a weaker peak structure, which we discuss separately below. Again saturation of $N_\xi$ {is not yet achieved for} {$N = 2^{17}$} in the low temperature case shown{, but for large $N=\{2^{32}, 2^{44}\}$ saturation is visible across the whole spectrum of eigenvalues}.

\begin{figure}
  \includegraphics[width=\linewidth]{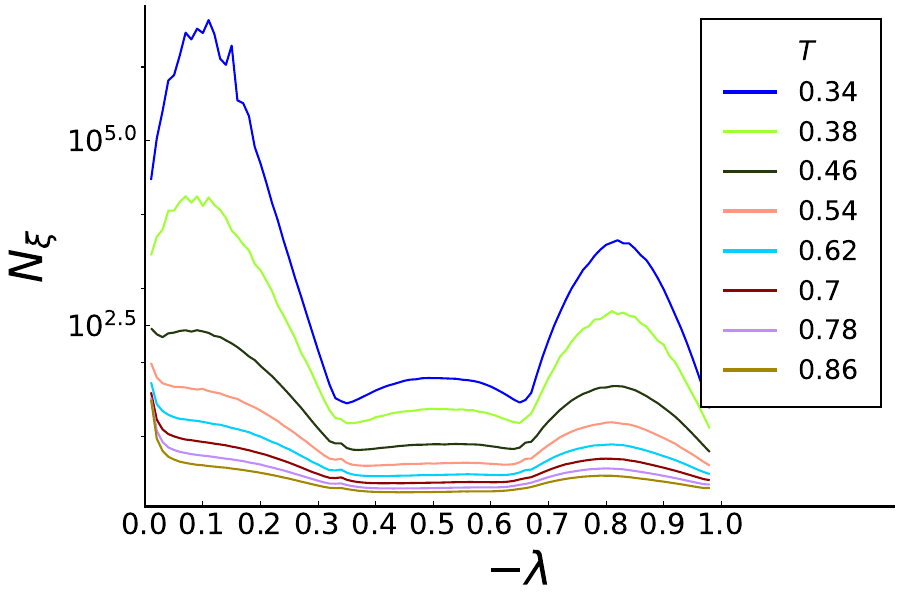}
  \caption{Correlation volume $N_\xi$ for  different temperatures in the glassy regime ($T<1$). Average over 10 independent runs. Effective network size $N = 2^{44}$. Population size $N_p = 10^5$. Eigenvalue grid size $\Delta \lambda = 0.01$}
        \label{allTs}  
 \end{figure}

In Fig.~\ref{allTs} we show the correlation volume $N_\xi$ for a whole range of temperatures below the glass transition -- since our estimates for $N_\xi$ generally exhibit statistical fluctuations that are smaller than those for $N I_2$, we concentrate on results for this quantity below. The key observation from Fig.~\ref{allTs} is that the non-monotonic variation across the spectrum continues to get stronger as $T$ is lowered. We remark that the figure displays results only for those temperatures for which we observe a saturation of $N_\xi$ at large $N$, with the corresponding limiting values being displayed in the plot. For temperatures lower than those shown, the estimator for $N_\xi$ does not saturate even for the highest effective size considered, $N = 2^{44}$, or becomes very noisy, especially in the low $|\lambda|$ region. We cannot therefore distinguish without further tests whether the correlation volume is finite and very large or in fact infinite, corresponding to slow modes that are genuinely localized. We leave this low temperature regime for future investigation.

{At this point, it is worth discussing how large we can make $N$, or correspondingly how small $\epsilon$, while producing meaningful results. In principle, nothing stops us from using $\epsilon$ of the order of the smallest floating point number, i.e.\ $\epsilon \sim \epsilon_0 = 10^{-300}$, and doing computations with this value. As discussed at the end of Sec.~\ref{sparsebm}, in the extended part of the spectrum and as long as $N_\xi$ is not too large, this is unproblematic because the results saturate already for $\epsilon\gg \epsilon_0$. In accordance with this expectation we find in this regime that the choice $\epsilon=\epsilon_0$ produces results indistinguishable from those for $N=2^{44}$ (data not shown).
Where states are localized or extended with a large correlation volume, on the other hand, one cannot directly use $\epsilon\sim \epsilon_0$ without checking independently whether the results saturate -- at a point depending on population size -- as $\epsilon $ is decreased from much larger values.}

\begin{figure}
  \centering 
  \begin{subfigure}[t]{0.4\textwidth}
    \caption{}
    \centering
        \includegraphics[width=\linewidth]{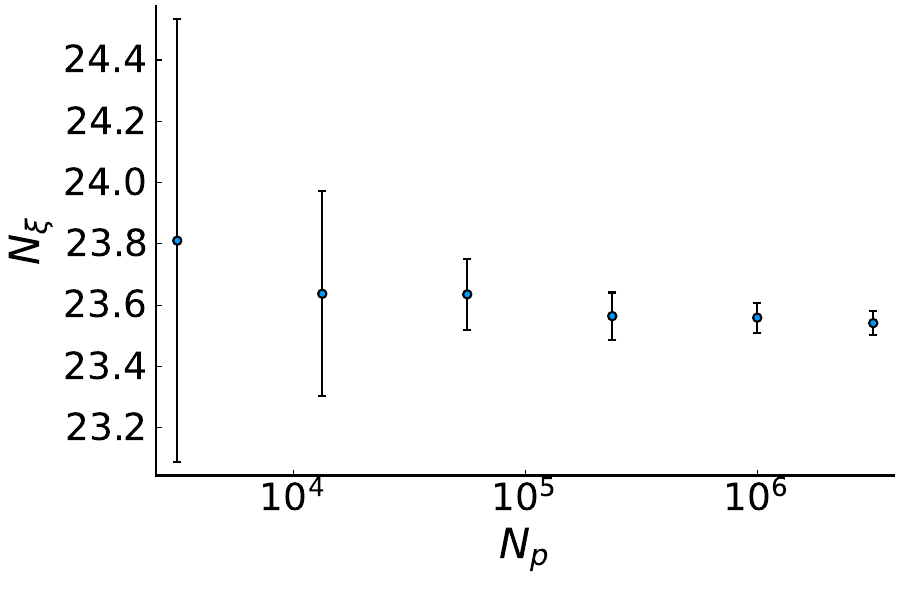} 
  \end{subfigure}
  \hfill
  \begin{subfigure}[t]{0.4\textwidth}
    \caption{}
    \centering
       \includegraphics[width=\linewidth]{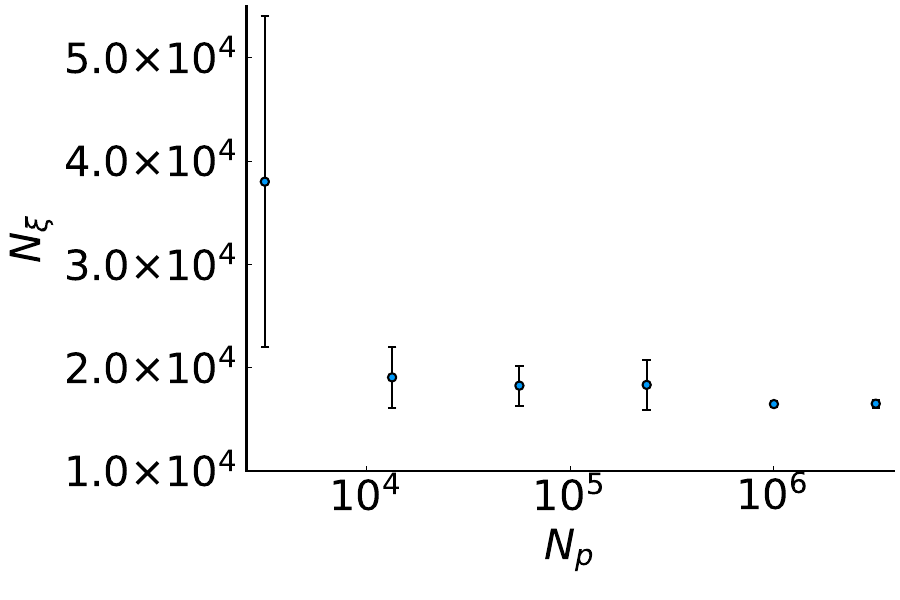} 
  \end{subfigure}
  \caption{Dependence of the correlation volume $N_\xi$ on population size $N_p$. $T = 0.38$, $\lambda = -0.5$ (top), $\lambda = -0.1$ (bottom). Effective network size $N = 2^{44}$. The error bars show half of the standard deviation obtained across 10 independent runs. }
  \label{scalingwithM}
\end{figure}

 Since our analysis relies on PD, we next discuss how the results depend on the population size. In Ref.~\cite{tikhonov2019critical} a systematic analysis of the dependence of $N_\xi$ on $N_p$  (or $M$ in the notation of~\cite{tikhonov2019critical}) is reported for the Anderson model, showing that above a certain critical $N_p^*$, $N_\xi$ becomes roughly independent of the population size.  According to the estimates in Ref.~\cite{tikhonov2019critical}, $N_p^* \sim N_\xi^{0.46}$. This scaling is expected to be  model dependent, and we do not aim to perform a detailed analysis of this point. However, the $N_p$-independence of $N_\xi$ for large $N_p$ is consistent with what we observe in our simulations. We show exemplary data demonstrating this in Fig.~\ref{scalingwithM}, for two different modes at the second lowest temperature in Fig.~\ref{allTs}.

\begin{figure}
  \includegraphics[width=\linewidth]{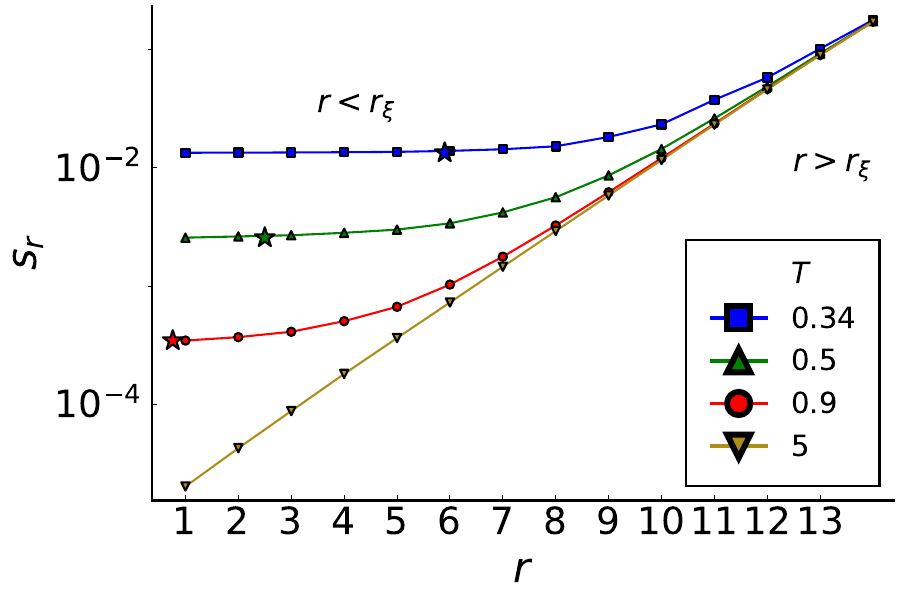} 
   \caption{Cumulative local DOS around the node with maximum value up to distance $r$ ($s_r$ from equation~\eqref{cumu}) for different temperatures and fixed $\lambda = -1/2$. Results are averages over 10 finite instances of size $N = 2^{18}$, using local values $\G_{ii}$ computed by SI. For comparison we also provide the estimated values of $r_\xi$ (converted from Figure~\ref{allTs} and {marked by stars}).}
   \label{internxi}
 \end{figure}

We now show the relevance of the correlation volume $N_\xi$ and the associated correlation length (eq.~\eqref{corrlength}) for finite network instances. We expect that within neighborhoods of size $N_\xi$ the local DOS is concentrated on a single node or a few nearby nodes, while within larger regions the local DOS should be spread out roughly uniformly. To test this for a given {network} and disorder realization, we first identify the node with maximum  $\G$, say $m = {\rm argmax} \G_{ii}$  and sum the contribution of its neighbours to the local DOS up to a distance $r$:
\begin{equation}
  s_r = \frac{1}{\sum_j \G_{jj}}\sum_{i: |i-m|\leq r}\G_{ii}
  \label{cumu}
\end{equation}
where the denominator normalizes the local DOS values to the analog of unit norm for a single eigenmode. Results for different temperatures at a fixed $\lambda$ are shown in Figure~\ref{internxi}. Below a distance of order $r_\xi$ we observe a plateau in the partial sums $s_r$ of the local DOS, indicating that most of the local DOS is concentrated in the maximum and its neighbours{, with the height of the plateau reflecting the extent of concentration}; for longer distances a growth away from the plateau is observed, as nodes further away start to make significant contributions to the accumulated DOS. In line with this, the plateau disappears at high $T$ in the fully extended regime, implying that all nodes contribute roughly equally to the cumulative local DOS.
 
\begin{figure}
  \centering 
  \begin{subfigure}[t]{0.4\textwidth}
    \caption{}
    \centering
    \includegraphics[width=\linewidth]{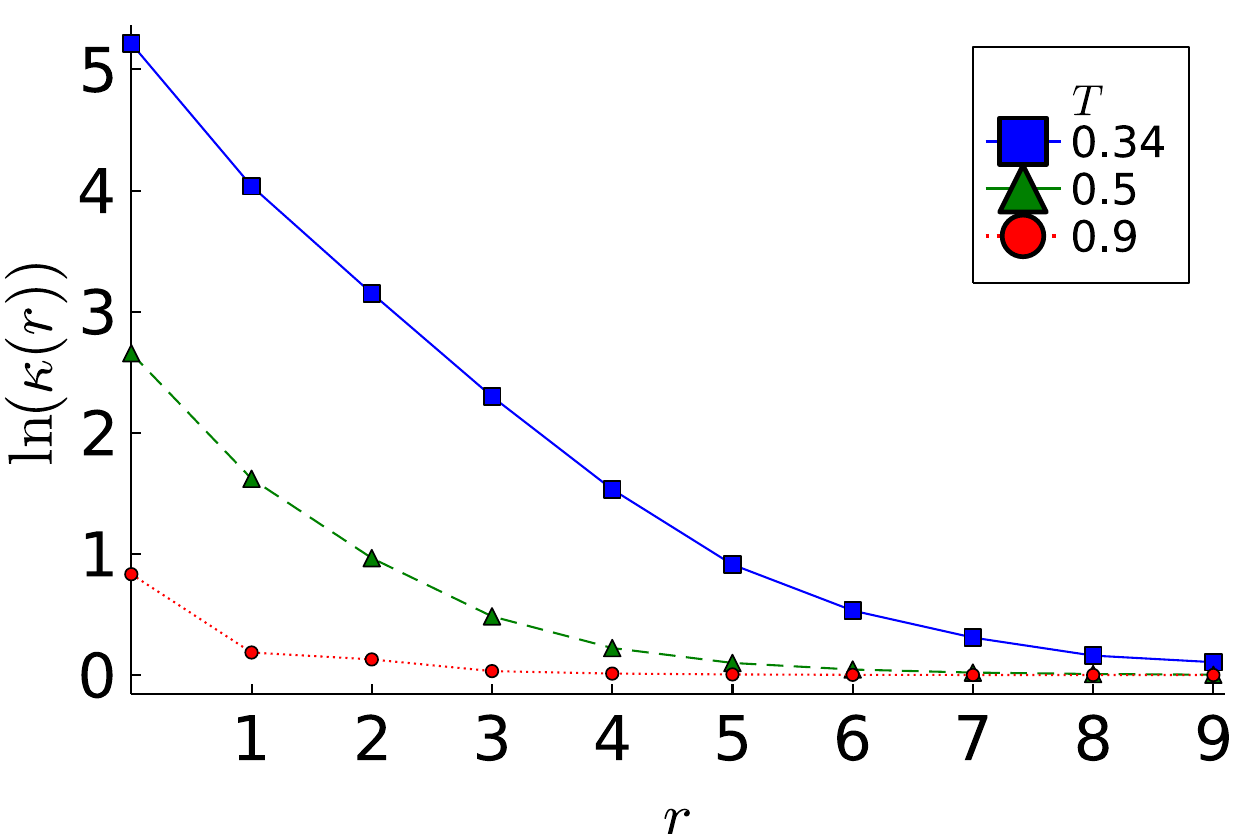}  
  \end{subfigure}
  \hfill
  \begin{subfigure}[t]{0.4\textwidth}
    \caption{}
    \centering
    \includegraphics[width=\linewidth]{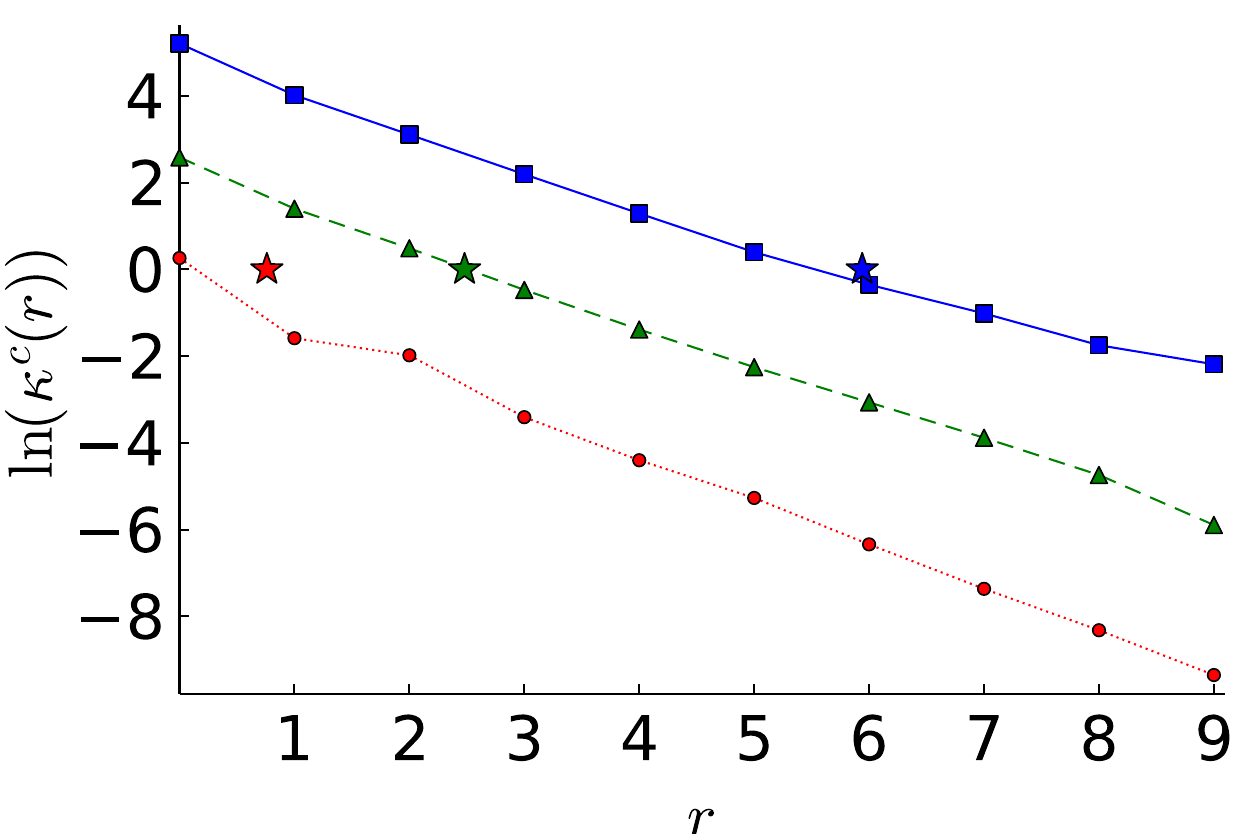} 
  \end{subfigure}
  \caption{Correlation function of the local DOS ((a), eq.~\eqref{corr1}) and connected correlation function ((b), eq.~\eqref{corr}) for different temperatures and $\lambda=-1/2$ on a single instance of size $N = 2^{19}$.  The stars indicate the correlation length $r_\xi$ estimated from PD. }
  \label{corrscaling}
\end{figure}

We can also use our data to interrogate the correspondence between the correlation volume and the correlation function introduced in eq.~\eqref{corr1}. For a large instance of size $N = 2^{19}$  we estimate $\kappa(r)$ and its connected version $\kappa^c(r)$ (eq.~\eqref{corr}) using SI for different temperatures. The result is shown in Fig.~\ref{corrscaling}.  For lengths $r \gg r_\xi$  we expect no correlation between the local DOS of two distant nodes, and therefore $\kappa \to 1$. Additionally, and following Ref.~\cite{tikhonov2019critical} {(where a similar correlation function is introduced, with the difference that it considers contributions from both imaginary and real parts of the resolvent),} we compare quantitatively the correlation length with the length scale extracted from the condition $\ln(\kappa^c(r)) = 0$. The plot in Fig.~\ref{corrscaling}(b) supports the idea that the comparison is meaningful and that $r_\xi$, which is a quantity extracted from the infinite--size (PD) analysis, is indeed relevant for describing correlations in finite networks.

We end this section with an analysis of localization properties around $\lambda = -1/3$, {where a peak in $NI_2$ is observed for low temperatures (Fig.~\ref{niprscaling})}; we have performed a similar investigation around the peak at $\lambda = -2/3$, with the same qualitative conclusions. As background we recall from previous work~\cite{tapias2020entropic} that the spectral density concentrates for $T \to 0$ on {delta} peaks around $\lambda^*_k = -k/c$ with $k \in \{0, \ldots, c\}$. These modes are associated with entropic effects: $k$ effectively labels the number of lower-lying neighbours of a node in question. By their nature, they are strongly localized (Appendix~\ref{locsec}). The question guiding us is then: how does the existence of these entropic zero temperature peaks affect the localization properties at $T>0$? 

\begin{figure}
   \includegraphics[width=\linewidth]{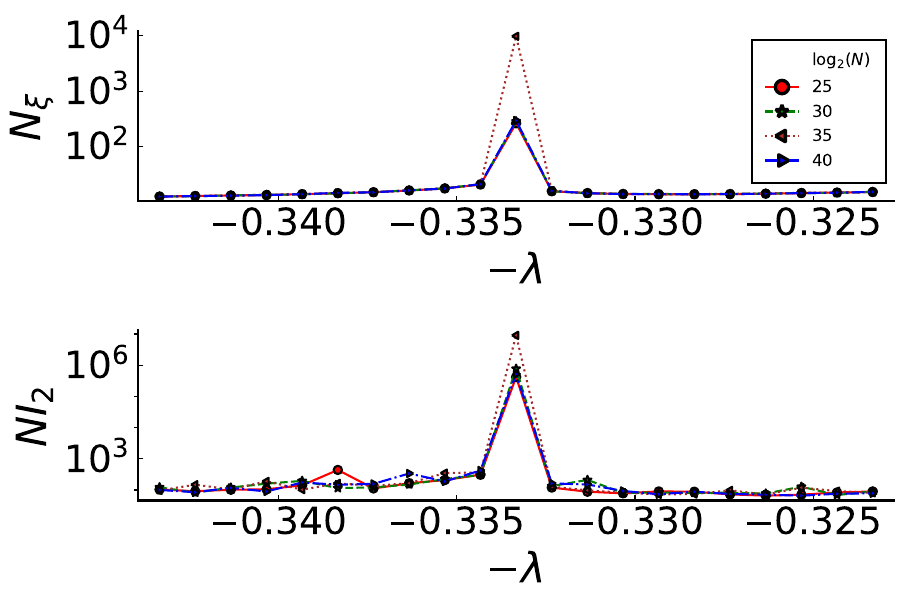} 
   \caption{Dependence of $N_\xi$ (top) and $N I_2$ (bottom) on effective network size $N$, for $T = 0.4$ in a small range around $\lambda = -1/3$. Population size $N_p = 10^5$, eigenvalue grid size $\Delta \lambda = 10^{-3}$.}
    \label{peaks}
  \end{figure}

\begin{figure}
  \centering 
  \begin{subfigure}[t]{0.4\textwidth}
    \caption{}
    \centering
    \includegraphics[width=\linewidth]{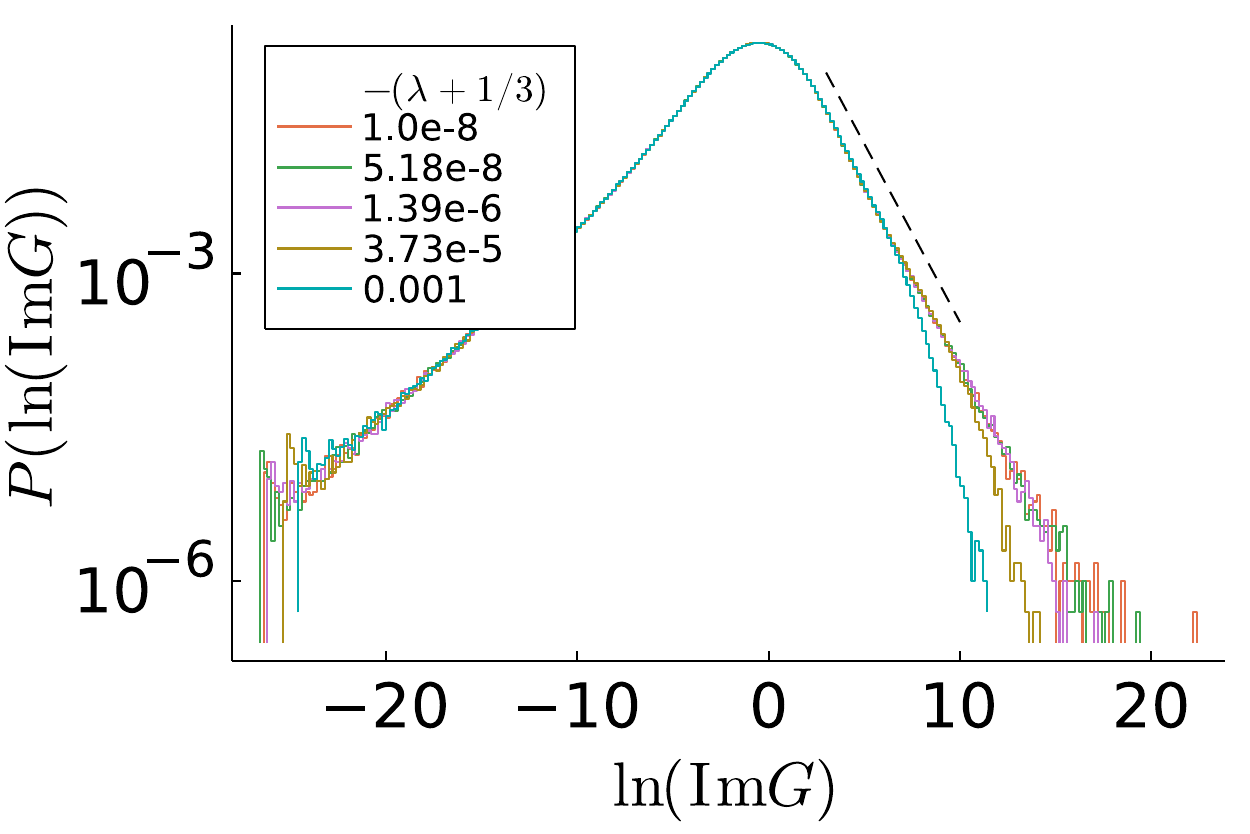}  
  \end{subfigure}
  \hfill
  \begin{subfigure}[t]{0.4\textwidth}
    \caption{}
    \centering
    \includegraphics[width=\linewidth]{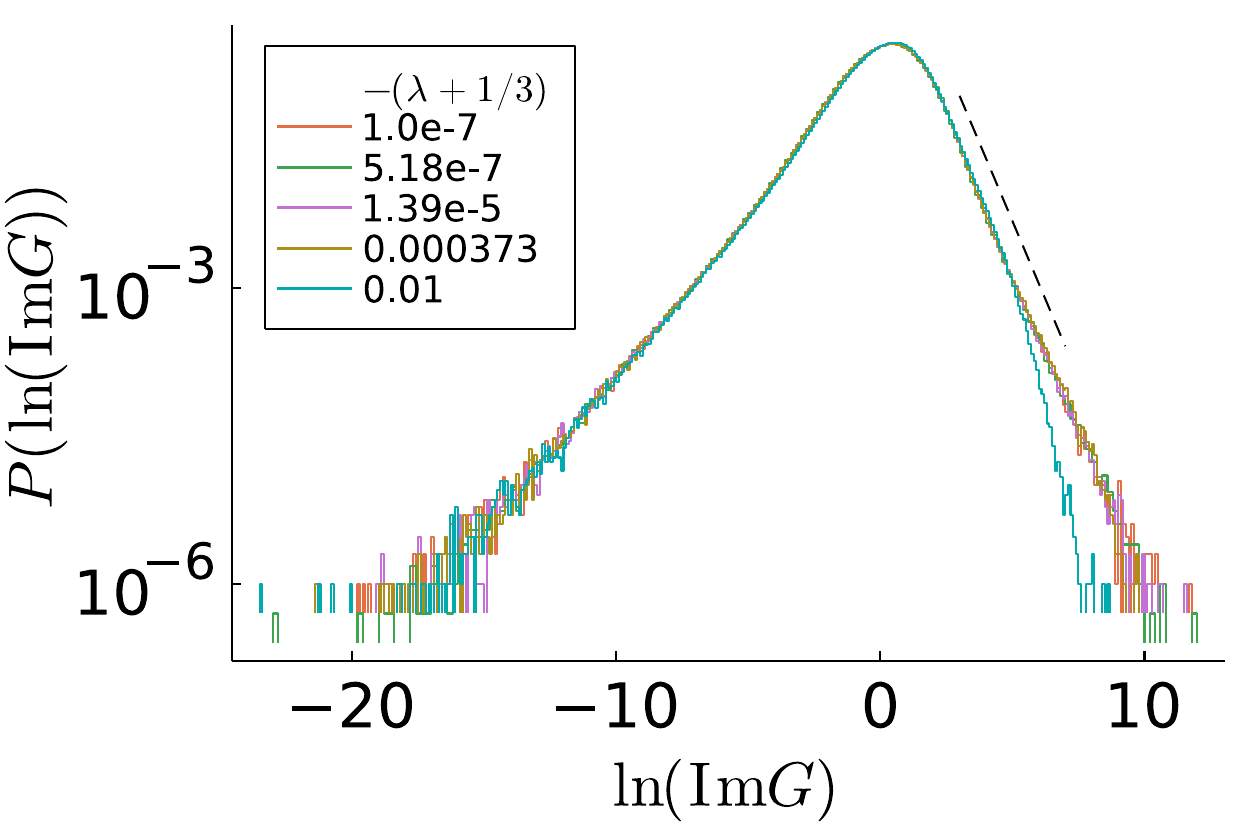} 
  \end{subfigure}
  \caption{Histograms of the full distribution $P(\G)$ (after a logarithmic transformation) for temperatures $T = 0.4$ (top) and $T = 0.7$ (bottom).  The dashed line shows the fit of the tail to $P(\G) \propto (\G)^{-\alpha} $ with $\alpha = 1.8$ (top) and $\alpha=2.46$ (bottom). Results are from PD with effective $N = 2^{40}$ ($\epsilon \sim 10^{-12})$ and population size $N_p = 10^5$, for values of $\lambda$ near the spectral peak as shown. }
  \label{fullimG}
\end{figure}

\begin{figure}
    \includegraphics[width=\linewidth]{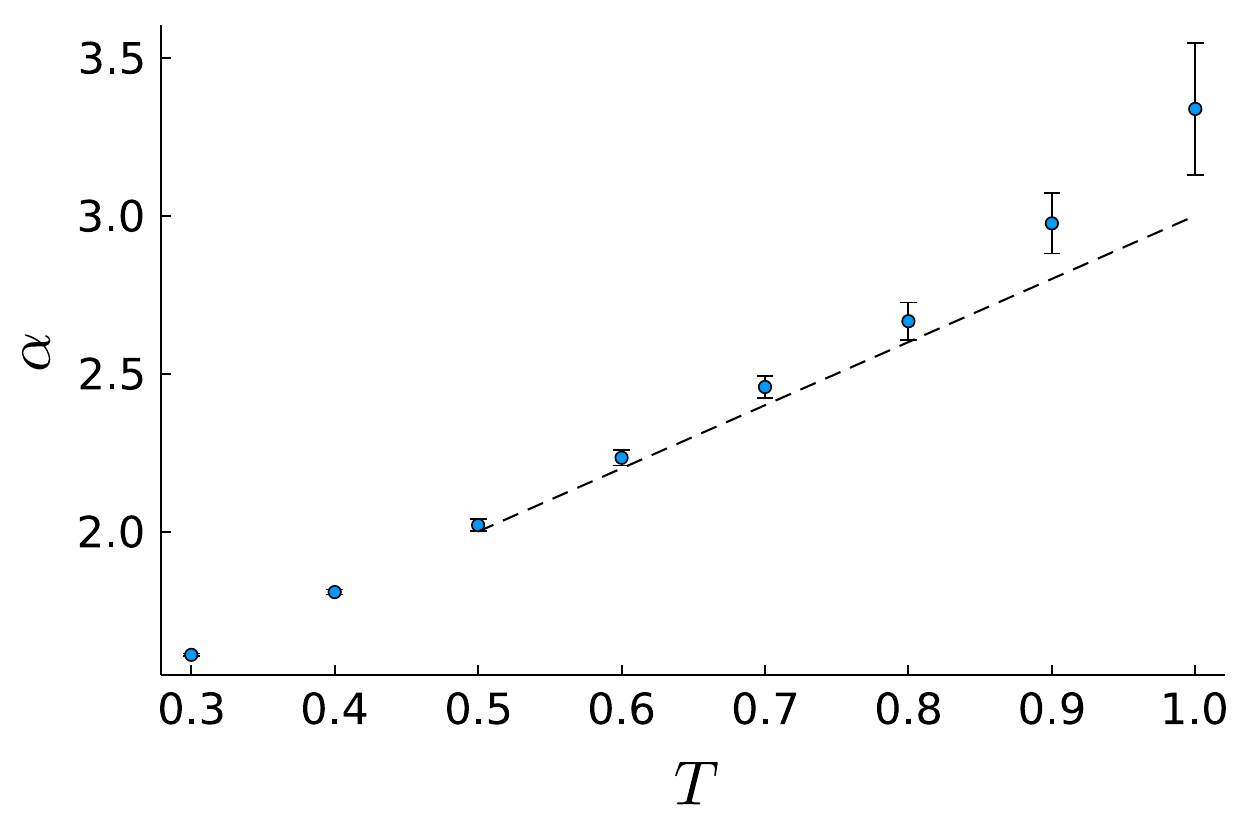} 
    \caption{Power law exponent of the right tail $P(\G) \propto \G^{-\alpha}$ at $\lambda = -1/3$. Exponents were fitted using the maximum likelihood estimator of Ref.~\cite{clauset2009power}; error bars are obtained by comparing estimates from different fitting intervals $(\G^{\rm{min}}, \G^{\rm{max}})$. {Dashed line: guide to the eye, interpolating linearly from $\alpha=2$ at $T=1/2$ to $\alpha=3$ at the glass transition temperature $T=1$.}}
    \label{alphaexp}
  \end{figure}

Figure~\ref{peaks} shows the scaling of $N_\xi$ and of the product $N I_2$ with the effective size $N$ -- thus within a PD calculation -- for a low temperature and in a narrow eigenvalue range around the peak. We observe that both $N_\xi$ and $NI_2$ are essentially $N$-independent for the large {values of} $N$ used here, {and thus the states} are extended eigenmodes. The exception is the {state for the} $\lambda$-value at the peak, where we see large (and apparently non-systematic) variations with $N$. To understand this we look more closely at the full distribution $P(\G)$ for the different values of $\lambda$ (Fig.~\ref{fullimG}).  One can see that this distribution develops a power law tail $P(\G) \sim (\G)^{-\alpha}$ for large $\G$, exactly as $\lambda$ approaches the peak value $\lambda^*_1$.  Repeating the same analysis for different temperatures, we obtain the exponent $\alpha$ as a function of $T$ (Fig.~\ref{alphaexp}). We observe that below $T < 1/2$, the exponent lies in the range $\alpha<2$: therefore already the first moment of $P(\G)$ diverges (in the limit $\epsilon \to 0$). This result is consistent with the observed divergence for the spectral density in the same regime~\cite{tapias2020entropic}. For the two highest temperatures $T=0.9$ and 1, just below and at the glass transition, our uncertainties in the estimation of $\alpha$ are largest. The other data points are consistent with a linear $T$-dependence of $\alpha$ that reaches the value $\alpha=3$ at $T=1$ {(dashed line in Fig.~\ref{alphaexp})}, with a transition to extended modes there. {Our numerical data do not, however, rule out that such a transition could happen already at a lower temperature around $T=0.9$.} {To resolve the question it would be worth pursuing an analytical calculation of the exact exponent of the power law as a function of $T$. This may be possible by a careful analysis of the cavity equations~\eqref{cavities} and is left for future investigation.}

The overall picture we gain from this part of our analysis is that for $T < 1/2$ the correlation volume diverges at the peak (because of a divergence of the first moment of $P(\G)$, but not of the typical value (cf.~Fig.~\ref{fullimG})) and that $N I_2$ also diverges. Thus, the localized nature of the modes at the special values $\lambda \in \{-1/3, -2/3\}$ remains for nonzero temperatures.  {We conjecture that this will be the case for all $\lambda^*_k$, i.e.\ all multiples of $-1/c$ in a generic sparse RRG}. In addition, for temperatures in the range $1/2 < T \lesssim 1$, the first moment of $P(\G)$ is finite, but the second moment is not. This implies that the correlation volume becomes constant while the product $N I_2$ diverges as $N$ grows, rather than saturating. This intriguing observation of the ``decoupling'' of the two different probes of localization could suggest the existence of extended non-ergodic states of multifractal structure in the relevant temperature regime. {Such states have been discussed extensively in the context of the Anderson model~\cite{de2014anderson, tikhonov2016fractality, kravtsov2018nonergodic, khaymovich2021dynamical}} (see also~\cite{monthus2017statistical}). Whether they exist for the BM model would therefore certainly require further investigation.

  \begin{figure}
\includegraphics[width=\linewidth]{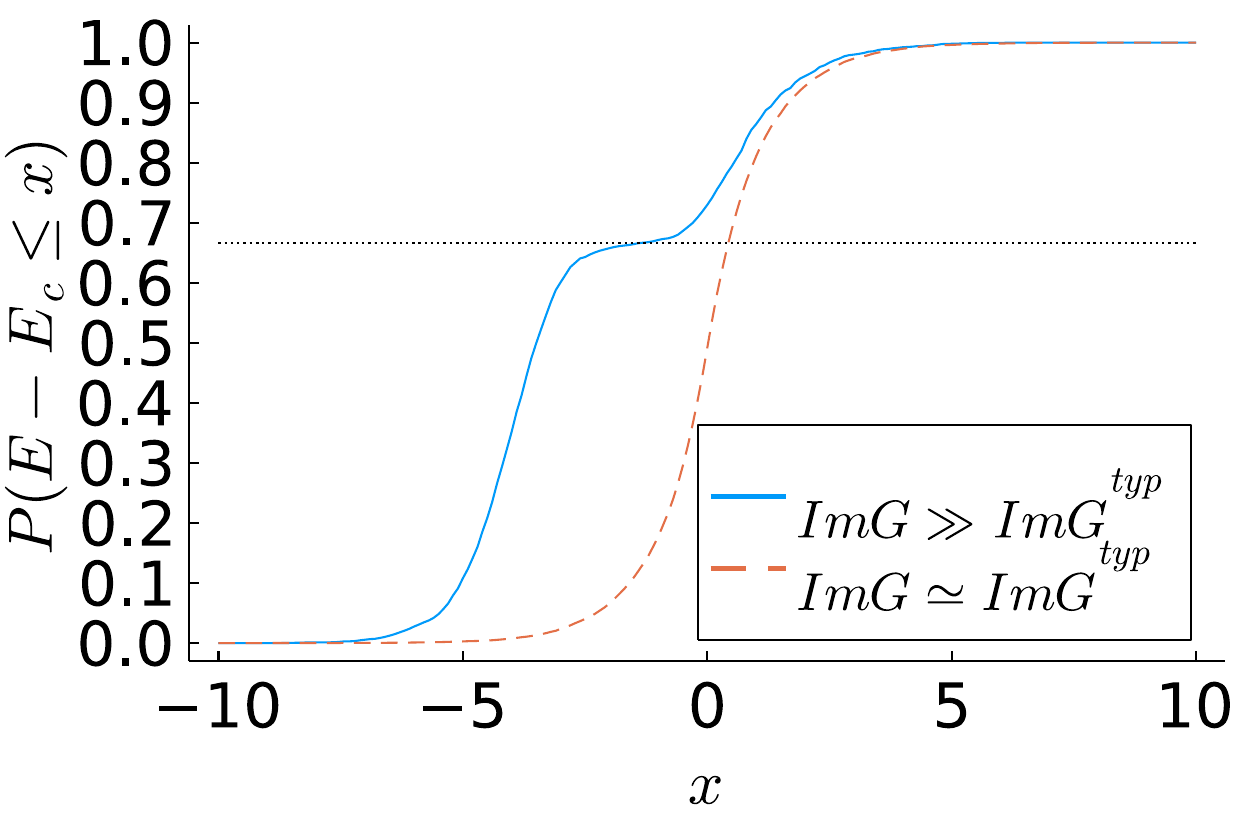} 
\caption{Cumulative distribution of the local trap depths $E$ around chosen network nodes of depth $E_c$ for $\lambda = -1/3$ and $T = 0.4$. Dashed line: typical nodes, solid line:  nodes with large $\G_{ii}$, i.e.\ in the tail of $P(\G)$ (see Fig.~\ref{fullimG}). Results from population dynamics with effective $N = 2^{40}$ and $N_p = 10^5$. {Dotted horizontal line: $P(E- E_c \leq x) = 2/3$, provided for comparison with the height of the observed plateau.}
}
\label{localconf}
\end{figure}

From the discussion so far we expect that the power law tail of $P(\G)$ at $\lambda = -1/3$ (and similarly at $\lambda = -2/3$) must be related to the entropic nature of the corresponding eigenmodes. To substantiate this, we compare (at a fixed low  temperature) nodes with typical values of $\G_{ii}$ to those with large $\G_{ii}$ in the tail of $P(\G)$. For each node we record its (``central'') trap depth $E_c$ and the trap depths $E$ of its neighboring nodes, and find the distribution of $E-E_c$.  The resulting cumulative distribution is shown in Fig.~\ref{localconf}.
{We observe} for the typical nodes that the trap depth difference $E-E_c$ is equally likely to be positive and negative, with no detectable structure in the distribution. For the nodes in the tail of $P(\G)$, on the other hand, the cumulative distribution shows a plateau at height $2/3$ and close to $E-E_c=0$: the probability for $E-E_c$ to be negative, i.e.\ for a neighbouring node to have smaller depth and thus be higher in the energy landscape  is $2/3$, which is exactly the result predicted from the entropic modes for $T\to 0$. This is additional evidence for the fact that the  localized modes we have found at nonzero temperature inherit the entropic structure of the $T=0$ modes.

\section{Conclusions}
\label{concs}

In this work, we have introduced a method to analyze {the localization properties of the states} of the Barrat--M\'ezard trap model defined on random regular graphs, based on a population dynamics approach. {We focused on the properties of the extended states as a function of the disorder parameter, which in our case is the (inverse) temperature.}

Our method has been inspired by earlier studies of the Anderson model on RRGs~\cite{abou1973selfconsistent, biroli2012difference, monthus2017statistical, biroli2018delocalization, tikhonov2019critical, tikhonov2019statistics}. It allows a systematic analysis of the extended eigenstates of any sparse system that may be solved using the cavity method, based on the properties of the distribution of the imaginary part of the diagonal resolvent entries $P(\G)$.

We showed that it is possible to obtain information on finite size instances by using the infinite--size self consistent equation for the cavity precisions, and used this approach to study the dependence of two localization probes -- inverse participation ratio and correlation volume -- on the (effective) system size. {A key point} of our method is the correspondence between the size of the system $N$ and the regularizer $\epsilon$ as given in equation~\ref{gene}. 

The advantages of population dynamics over the single instance cavity method are twofold: (i) the estimation of the full distribution $P(\G)$ for a given $\epsilon$ improves while requiring {less computational effort} (see Fig.~\ref{sivspd}) and (ii) it is possible to predict the behavior of very large instances as the population size can generically be smaller than the instance size. Indeed, we have considered effective system sizes up to $N = 2^{44}$, which would be impossible to analyze even using the most efficient currently available numerical routines for finite {networks}. {This last point is very relevant because it implies that we can estimate with improved accuracy the correlation volume close to the localization transition, and obtain better data to estimate e.g.\ the functional form of the divergence there. This will be left for future work.}

An important outcome of our work is that we were able to estimate the mean inverse participation ratio of finite size instances quantitatively, and not only its scaling with system size as is usually done. The relevant estimator is given in equation~\eqref{iprtikh}, to be used in conjunction with the rule~\eqref{gene}.  Equally importantly, we have given a more concrete interpretation for the correlation volume $N_\xi$ in terms of the contribution to the local density of states: within a neighbourhood of size $N_\xi$ around the node with maximum local DOS, most nodes make only a negligible contribution to the cumulative local DOS so that -- within this region -- the system looks as if it were fully localized.

Turning to the specific results for the BM model, {we highlight the observed non-monotonic, highly non-trivial variation of $N_\xi$ and $N I_2$ across the spectrum. For a given temperature and connectivity $c =3$, we can distinguish at least four different regions in the bulk of the spectrum with a distinct tendency towards localization. Referring to Fig.~\ref{allTs}, we have firstly the central region between the peaks, $-2/3<\lambda<-1/3$, with the lowest correlation volumes and so the most delocalized states. Secondly, the region $|\lambda| < 1/3$ exhibits the highest correlation volumes, which occur for states  associated with activation processes~\cite{tapias2020entropic}. Thirdly, the region of high relaxation rates $|\lambda| > 2/3$ has moderately extended modes. Their properties are tied more to the structure of the network than to the disordered energy landscape; see for instance Ref.~\cite{margiotta2018spectral} for confirmation of this statement at the level of the density of states of the Bouchaud trap model. Finally, the point sets $\lambda \in \{-1/3, -2/3 \}$ have localized eigenmodes for low enough temperatures; we discuss these separately below.
Notice that we left out of our analysis the mobility edges near $\lambda=0$ and $\lambda=-1$, where there is a more conventional transition to localized states (see Ref.~\cite{margiotta2018spectral} for an analysis of this in the Bouchaud trap model, and for instance Ref.~\cite{biroli2012difference} for a characterization in the Anderson model).}

A characteristic feature of the BM model is the existence of localized modes at $T = 0$ that physically represent entropic effects.  {In this limit, each mode with associated eigenvalue $-k/c$ corresponds to a local configuration in which the total probability mass is concentrated in a single node with $k$ neighbours that lie at greater depth in the energy landscape. Thus, the origin of localization is linked strongly to the energetic disorder here, rather than merely to the structure of the network.} {This mechanism is then quite different from what is observed on Erd\"os--R\'enyi graphs, where there are also sets of localized states at discrete values of $\lambda$: the properties of these states depend entirely on the structure of the network. As a matter of fact, the associated eigenvalues correspond to those of the adjacency matrix of finite trees, and the localization effects can be explained in terms of ``dangling trees'' embedded in the giant component~\cite{bauer2001random, golinelli2003statistics, kuhn2016disentangling}.} 
{In terms of the spectral properties, the situation in the BM model is similar to the Erd\"os--R\'enyi case with Gaussian couplings~\cite{kuhn2008spectra}, where a power law divergence of the DOS  occurs in the center of the spectrum, with associated localized eigenmodes. Nevertheless, as argued above the physics is rather different in our case.}

We have rationalized the localization of the singular ``entropic'' modes from the existence of a power law tail of the distribution $P(\G)$, which causes a divergence of the first or higher moments. This has similarities to the Anderson model, where it is known that for localized states $P(\G) \sim \G^{-\beta}$ for $\G\gg 1$ with $\beta \leq 3/2$ (see Refs.~\cite{abou1973selfconsistent, ciliberti2005anderson}). However, our exponents are different (see Fig.~\ref{alphaexp}) and the typical value of $P(\G)$ remains $O(1)$ for $\epsilon \to 0$ in our case {in contrast to the Anderson model~\cite{kravtsov2018nonergodic}}. Both observations point to the fact that the physical mechanisms for localization are distinct in the BM model.

In future work it {will be worth investigating the phase diagram of the model studied here down to $T=0$: as in the zero temperature limit all eigenstates are fully localized, one might expect that the isolated localized modes at $\lambda=-k/c$ that we found broaden into multiple localized regions of the spectrum at lower $T$ than studied here.} {The application of our method to other disordered systems for which the cavity method is valid will also be interesting to explore, including e.g.\ biological networks defined on linear chains~\cite{amir2016nonhermitian, tanaka2019nonhermitian}.} {From a broader perspective, we know that localization plays an important role in the dynamical properties of disordered systems~\cite{biroli2017delocalized, de2020subdiffusion, tikhonov2021anderson}. This role will be fascinating to elucidate also in sparse trap models in the presence of a dynamical bias, within the context of trajectory thermodynamics~\cite{garrahan2009first, jack2010large, jack2015effective, jack2020ergodicity}}.

\section*{Acknowledgments}

We thank Reimer K\"uhn, Fernando Metz, Federico Ricci-Tersenghi and Marco Tarzia for valuable discussions and feedback.

\bibliographystyle{unsrt}
\bibliography{cavi}

\section{Appendix}
 
\setcounter{figure}{0} \renewcommand{\thefigure}{A.\arabic{figure}}

\subsection{Agreement between Exact Diagonalization and Single Instance cavity method}
\label{agreement}

\begin{figure}
  \centering 
  \begin{subfigure}[t]{0.4\textwidth}
    \caption{}
    \centering
    \includegraphics[width=\linewidth]{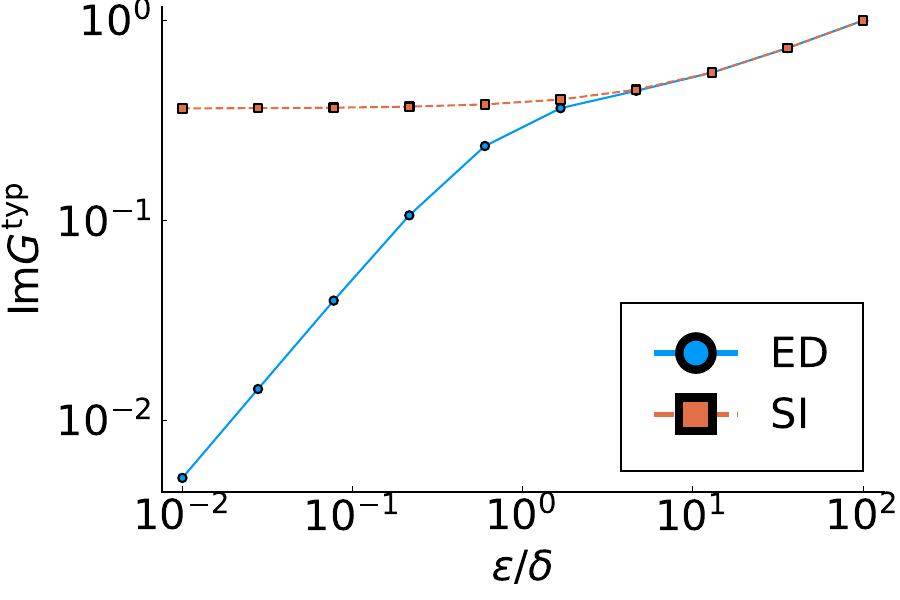}  
  \end{subfigure}
  \hfill
  \begin{subfigure}[t]{0.4\textwidth}
    \caption{}
    \centering
    \includegraphics[width=\linewidth]{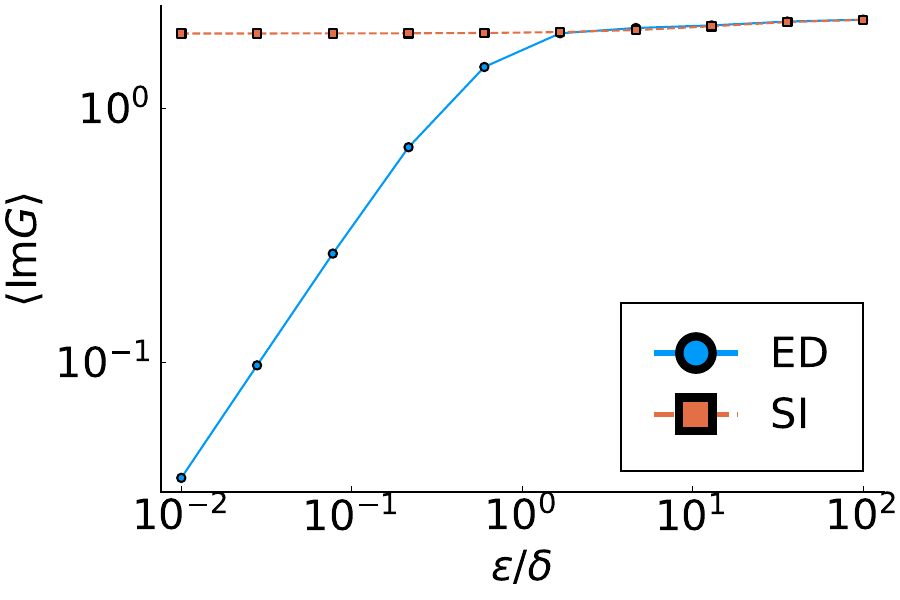} 
  \end{subfigure}
  \caption{Typical (top) and mean (bottom) values of $P(\G)$ for $T = 0.5$ in the middle of the spectrum ($\lambda = -1/2$), estimated with Exact Diagonalization (ED) and Single Instance (SI) cavity method, as a function of $\epsilon/\delta$ with $\delta = 1/(\rho(\lambda) N)$ the mean level spacing. Size of the instance: $N = 2^{14}$ }
  \label{typ_mean}
\end{figure}

Here we show the agreement between Exact Diagonalization (ED) and the Single Instance (SI) cavity method in a certain interval of $\epsilon$ around the mean level spacing.  We do this by comparing the mean $\langle\G\rangle$ and the typical value $\G^{\rm typ}={\rm{e}}^{(\langle\ln(\G)\rangle)}$ of $\G$. A generic case is shown in Fig.~\ref{typ_mean}. By ED we refer to the direct calculation of the resolvent entries $\G_{ii}$ for a given instance in terms of the eigenvectors $\bm{v}_\alpha$ and eigenvalues $\lambda_\alpha$ of the (symmetrized) master operator as:
\begin{align}
  {\bm{G}}(\lambda-i\epsilon) = \sum_{\alpha = 0}^{N-1} \frac{\v_\alpha \v_\alpha^T}{\lambda - i \epsilon - \lambda_\alpha}
  \label{resmatrixexp}
\end{align}
This implies
\begin{align}
  \G_{ii} = \epsilon \sum_{\alpha = 0}^{N-1} \frac{v_{\alpha,i}^2}{(\lambda- \lambda_\alpha)^2 + \epsilon^2}
  \label{ImG}
\end{align}
In Fig.~\ref{typ_mean} one notices that there is a regime (which increases with $N$ as shown in Ref.~\cite{biroli2018delocalization}) around  $\epsilon/\delta = 1$ (where $\delta=1/(\rho(\lambda)N)$ is the mean level spacing) that satisfies two conditions: (i) the results between SI and ED agree or are very close, and (ii) they are roughly  $\epsilon$ independent. Any prefactor $C/\pi$ of order unity in $\epsilon = C/(\pi \rho(\lambda) N)$ within this regime is equally valid. 

\subsection{IPR estimation}
\label{iprest}

In this section, we will show the superiority of the IPR estimator (eq.~\eqref{iprtikh}) in comparison with the standard estimator used in conjunction with population dynamics~\cite{mirlin2000statistics, metz2010localization, biroli2012difference} (see equation~\eqref{bollem} below). We will do this in two ways. First, by showing numerical results that compare both estimators with direct diagonalization. Second, by computing the resolvent entries analytically in the $T \to \infty$ limit where the entries of the eigenvectors in the bulk become Gaussian distributed~\cite{clark2018moments, backhausz2019almost}.

\subsubsection{Metz, Neri, Boll\'e estimator}

The common estimator for the IPR in terms of the Green's functions rederived by Metz, Neri and Boll\'e~\cite{metz2010localization} and originally used in the context of supersymmetry theory~\cite{mirlin2000statistics} is
 \begin{align}
   \hat{I}_2 (\lambda) = \frac{1}{\pi \rho} \lim_{\epsilon \to 0} \epsilon \langle |G|^2 \rangle
   \label{ibolle}
 \end{align}
 The usual approach to studying localization with this formula is to look at the scaling of $\hat{I}_2$ with decreasing $\epsilon$; if it converges to a value that is independent of $\epsilon$, the states are localized, whereas if the IPR scales as $\sim \epsilon$, the states are extended. This makes sense through the lens of the prescription~\eqref{gene}, as decreasing $\epsilon$ corresponds to increasing $N$.

 However, \emph{per se} formula~\eqref{ibolle} cannot predict the mean IPR of instances of size $N$ because it is intended to work in the thermodynamic limit. To address this, we relax the limit and substitute $\epsilon$ from eq.~\eqref{gene} to get
 \begin{align}
   \hat{I}_2 &= \frac{C}{\pi^2 \rho^2 N}  \langle |G|^2 \rangle \\
   &= \frac{C}{N}  \frac{\langle |G|^2 \rangle }{\langle \G \rangle^2}
   \label{bollem}
 \end{align}

 \subsubsection{Numerical comparison}

\begin{figure}
   \begin{subfigure}[t]{0.4\textwidth}
    \caption{}
    \centering
    \includegraphics[width=\linewidth]{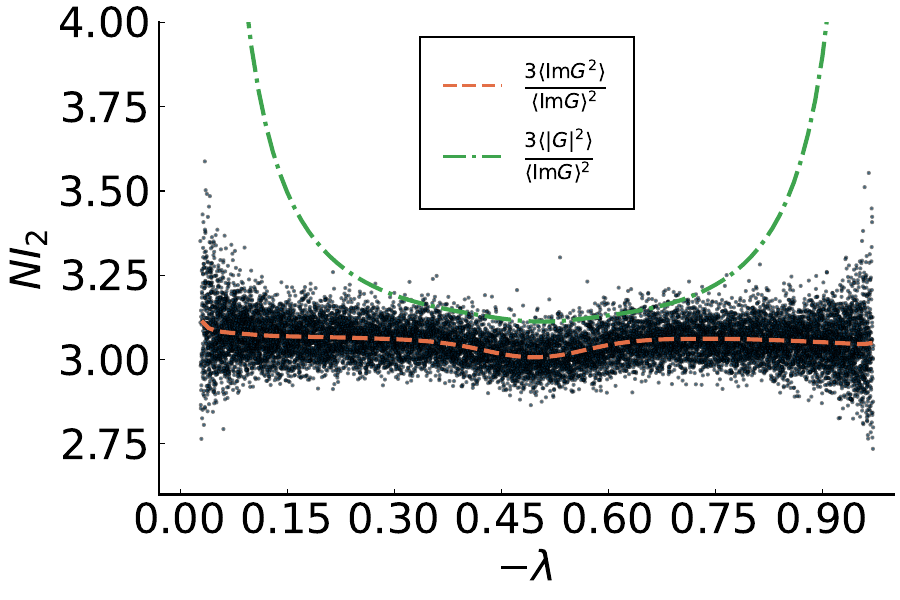} 
  \end{subfigure}
  \hfill
  \begin{subfigure}[t]{0.4\textwidth}
    \caption{}
    \centering
    \includegraphics[width=\linewidth]{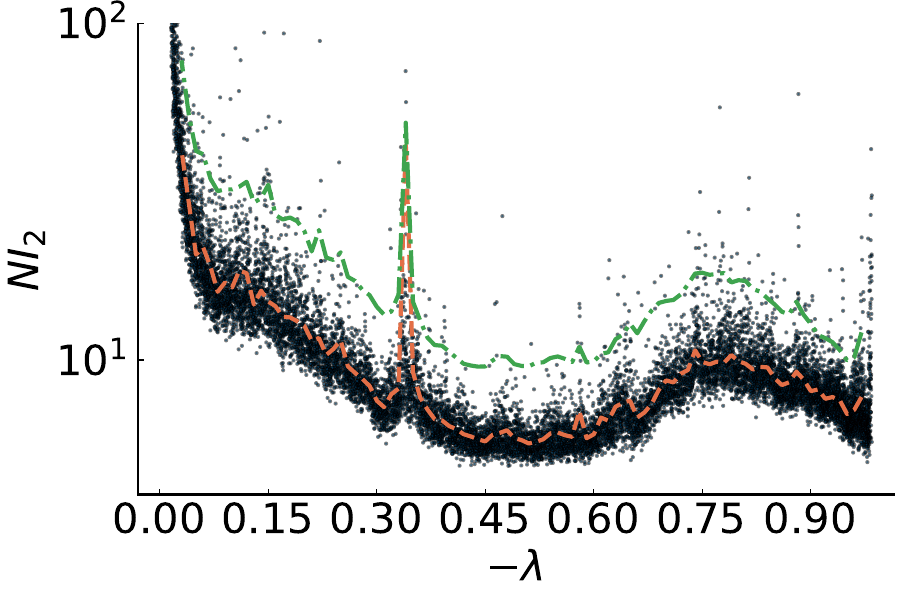} 
  \end{subfigure}
  \hfill
  \begin{subfigure}[t]{0.4\textwidth}
    \caption{}
    \centering
    \includegraphics[width=\linewidth]{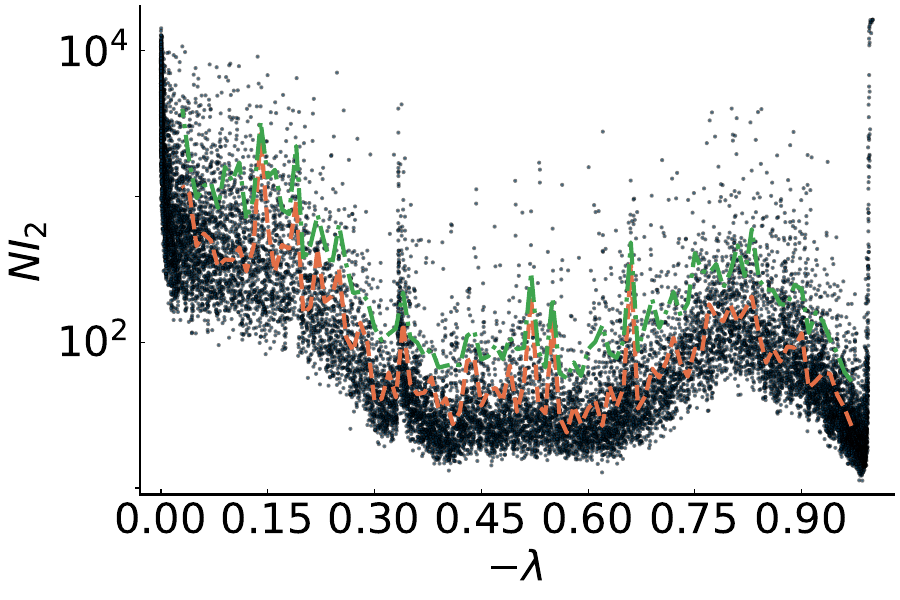} 
  \end{subfigure}
   \caption{Inverse participation ratio (multiplied by $N$) for three different single instances of the master operator. Size of the network $N=2^{14}$, temperatures $T = 10$ (top), $T = 1$ (middle), $T = 0.5$ (bottom). Scatter plot: direct diagonalization (eq.~\eqref{empiripr}). Dashed line: estimator~\eqref{iprtikh}, dash-dotted line: estimator~\eqref{bollem},  both obtained using the single instance cavity method (eqns.~\eqref{cavities}--\eqref{marginals}) with eigenvalue grid size $\Delta \lambda = 0.01$.}
    \label{manyTs}
  \end{figure}

 In Fig.~\ref{manyTs} we compare the estimator~\eqref{iprtikh} with~\eqref{bollem} for different temperatures. For a simple comparison we fixed the constant in equation~\eqref{bollem} as $C = 3$. 
  
  As the figure reveals, the estimator~\eqref{iprtikh} fits the results from direct diagonalization better, especially in the fully extended regime (high $T$). We show in the next subsection how this can be rationalized. From the figure, one can also observe that for moderate disorder, the estimator~\eqref{bollem} is qualitatively correct and differs from the actual result by a disorder--dependent prefactor. If one is only concerned with the scaling of the IPR with $\epsilon$, then both estimators are equally good in this regime.

\subsubsection{High $T$ limit}

Here, we will start from the definition of the resolvent in terms of the eigensystem (eq.~\eqref{ImG}) to evaluate the IPR estimators \eqref{iprtikh} and~\eqref{bollem} analytically in the limit $T \to \infty$, where the entries of the eigenvectors become Gaussian distributed and the density of states follows the Kesten--McKay law~\cite{kesten1959symmetric, mckay1981expected}.

Let us consider the definition of the density of states (eq.~\eqref{sidos}) and rewrite the estimator~\eqref{iprtikh} as
  \begin{align}
    I_2(\lambda)  &= \frac{3}{N \pi^2 \rho(\lambda)^2} \langle \G^2 \rangle \\
    &= \frac{3}{N^2 \pi^2 \rho(\lambda)^2} \sum_{i=1}^N  \G_{ii}^2 
  \end{align}
  Then we replace the resolvent entries by eq.~\eqref{ImG}, which gives
  \begin{align}
    I_2(\lambda)  = \frac{3}{N^2 \pi^2 \rho(\lambda)^2} \sum_{i=1}^N  \sum_{\alpha, \gamma} \frac{ \epsilon^2 v_{\alpha,i}^2 v_{\gamma,i}^2}{( (\lambda- \lambda_\alpha)^2 + \epsilon^2) ((\lambda- \lambda_\gamma)^2 + \epsilon^2)}
    \label{i21}
  \end{align}
  We then consider that the eigenvector components are Gaussian distributed with mean zero and variance $1/N$  and also that any two different eigenvectors are uncorrelated. These considerations give the following auxiliary relation for large $N$:
  \begin{align}
    \frac{1}{N} \sum_{i=1}^N  {v_{\alpha,i}^2 v_{\gamma,i}^2} &= \langle {v_{\alpha,i}^2 v_{\gamma,i}^2}\rangle \\
   &= \frac{1}{N^2} (1 + 2 \delta_{\alpha, \gamma})
\label{gauss}
  \end{align}
since $\langle {v_{\alpha,i}^2 v_{\gamma,i}^2}\rangle=
\langle {v_{\alpha,i}^2\rangle \langle v_{\gamma,i}^2}\rangle=(1/N)^2$ for $\alpha\neq \gamma$, but $=3(1/N)^2$ for $\alpha=\gamma$.
Substitution of this result into equation~\eqref{i21} yields
  \begin{align}
    I_2(\lambda)  = \frac{3}{N^3 \pi^2 \rho(\lambda)^2} \sum_{\alpha, \gamma} \frac{ \epsilon^2 (1 + 2  \delta_{\alpha, \gamma})}{( (\lambda- \lambda_\alpha)^2 + \epsilon^2) ((\lambda- \lambda_\gamma)^2 + \epsilon^2)}
  \end{align}
Without the Kronecker delta term, the sum divided by $N^2$ would just be 
$\pi^2 \rho(\lambda)^2$, bearing in mind that in the definition of $\rho(\lambda)$ Lorentzian functions are used to represent Dirac deltas (see for instance, Ref.~\cite{margiotta2018spectral}). The previous expression thus simplifies to
  \begin{align}
    I_2(\lambda)  = \frac{3}{N \pi^2 \rho(\lambda)^2} \left( \pi^2 \rho(\lambda)^2 +  \frac{1}{N^2} \sum_{\alpha} \frac{ 2 \epsilon^2  }{ ( (\lambda- \lambda_\alpha)^2 + \epsilon^2)^2}\right)
    \label{mm}
  \end{align}
  The last term in eq.~\eqref{mm} can be seen as a representation of a Dirac delta because
  \begin{align}
    \frac{ 2 \epsilon^2  }{ ( (\lambda- \lambda_\alpha)^2 + \epsilon^2)^2} &= \frac{2}{\epsilon^2} \frac{1}{ (( (\lambda- \lambda_\alpha)/\epsilon)^2 + 1)^2} \\
                                                                           &= \frac{\pi}{\epsilon^2} f\left(\frac{\lambda- \lambda_\alpha}{\epsilon} \right) 
  \end{align}
  where $f(x) = 2/\pi (1/(1+x^2)^2) $ is a function that integrates to 1. The function $f(x)/\epsilon$ is thus a ``nascent'' delta function. Plugging this into~\eqref{mm} leads to
  \begin{align}
    I_2(\lambda)  &= \frac{3}{N \pi^2 \rho(\lambda)^2} \left( \pi^2 \rho(\lambda)^2 +  \frac{\pi}{N^2 \epsilon} \sum_\alpha \delta(\lambda - \lambda_\alpha) \right)\\
                  &= \frac{3}{N \pi^2 \rho(\lambda)^2} \left( \pi^2 \rho(\lambda)^2 +  \frac{\pi}{N \epsilon} \int\rho(\tilde{\lambda}) \delta(\lambda - \tilde{\lambda})  d \tilde{\lambda}  \right) \\
    &= \frac{3}{N} \left(1 + \frac{1}{N \pi \rho(\lambda) \epsilon }\right)
  \end{align}
  Finally, using our convention for $\epsilon$ (eq.~\eqref{gene}) we obtain
  \begin{align}
    I_2(\lambda)  = \frac{3}{N} \left(1 + \frac{1}{C} \right)
    \label{ex1}
  \end{align}
  This result is the prediction of the estimator~\eqref{iprtikh} in conjunction with~\eqref{gene} when the resolvent is constructed by exact diagonalization. Note that for our choice $C = 3$ this leads to the result $I_2(\lambda) = 4/N$, which deviates from the theoretical expectation and also from what is observed with direct diagonalization and from population dynamics (or single instance) results, see Fig.~\ref{manyTs} (top). The explanation behind this is that in contrast to the estimation from PD or SI where the resolvent distribution $P(\G)$  becomes $\epsilon$ independent for $N$ sufficiently large if the state is extended, the estimation by ED of the whole distribution is always $\epsilon$ dependent (even in the $\epsilon$ window where the mean and the typical value are constant), and the relative fluctuations of the local DOS captured by our IPR estimator reflect this dependence. Nevertheless, the analytical result~\eqref{ex1} illustrates that the  estimator~\eqref{i21} in this high $T$ limit does not depend on $\lambda$, therefore it is constant once $\epsilon$ is fixed and thus qualitatively correct (cf.~Figure~\ref{manyTs} (top)).

  Next we perform a similar analysis for the estimator~\eqref{bollem}. It is easy to see that this estimator contains~\eqref{ex1} but with a different prefactor plus the contributions from the real part of the resolvent (eq.~\eqref{resmatrixexp}). This gives
  \begin{align}
    \hat{I}_2(\lambda) = \frac{C}{N} \left(1 + \frac{1}{C} \right) +  \frac{C}{N^2 \pi^2 \rho(\lambda)^2} \times  \\
  \times  \bigg( \sum_{i=1}^N  \sum_{\alpha, \gamma} \frac{ v_{\alpha,i}^2 v_{\gamma,i}^2 (\lambda - \lambda_\alpha) (\lambda- \lambda_\gamma)}{( (\lambda- \lambda_\alpha)^2 + \epsilon^2) ((\lambda- \lambda_\gamma)^2 + \epsilon^2)} \bigg)
    \label{i2sim}
  \end{align}
  We now use again the Gaussanity of the eigenvector entries (eq.~\eqref{gauss}) and separate the last term in brackets into the term from {the Kronecker delta and the rest.}
  The latter contribution is
  \begin{align}
    \frac{1}{N} \left( \sum_\alpha  \frac{(\lambda - \lambda_\alpha)}{ (\lambda- \lambda_\alpha)^2 + \epsilon^2)} \right)^2 &= N \left( \int  d\tilde{\lambda} \frac{\rho(\tilde{\lambda}) (\lambda - \tilde{\lambda}) }{( (\lambda- \tilde{\lambda})^2 + \epsilon^2)} \right)^2 \\
    &=N \left( {\rm{PV}} \bigg(\frac{1}{\lambda - \tilde{\lambda}} \bigg)  \rho(\tilde{\lambda}) \right)^2
  \end{align}
  where $\rm{PV}$ refers to the Cauchy principal value. The Kronecker delta contribution, on the other hand,
  is
  \begin{align}
    \frac{2}{N} \sum_\alpha  \frac{(\lambda - \lambda_\alpha)^2}{ ((\lambda- \lambda_\alpha)^2 + \epsilon^2)^2} &= \frac{2}{N \epsilon^2} \sum_\alpha  \frac{((\lambda - \lambda_\alpha)/\epsilon)^2}{ (((\lambda- \lambda_\alpha)/\epsilon)^2 + 1)^2} \\
    &= \frac{\pi}{N \epsilon^2} \sum_\alpha g\left( \frac{\lambda - \lambda_\alpha}{\epsilon} \right) 
  \end{align}
  where $g(x) = 2/\pi (x^2/(1 + x^2)^2)$ is a function that integrates to 1. Again, the function $g(x)/\epsilon$ is  a ``nascent'' delta function and this contribution simplifies to
  \begin{align}
    \frac{\pi}{N \epsilon^2} \sum_\alpha g\left( \frac{\lambda - \lambda_\alpha}{\epsilon} \right)  =   \frac{\pi \rho(\lambda)}{\epsilon} 
  \end{align}
  Putting all the contributions together into~\eqref{i2sim} and replacing $\epsilon$ according to our convention~\eqref{gene} we get
  \begin{align}
     \hat{I}_2(\lambda) &= \frac{C}{N} \left(1 + \frac{2}{C} \right) +  \frac{C}{N \pi^2 \rho(\lambda)^2}  \left( {\rm{PV}} \bigg(\frac{1}{\lambda - \tilde{\lambda}} \bigg)  \rho(\tilde{\lambda}) \right)^2
  \end{align}
So far the derivation applies to a generic DOS. For concrete results, we take this spectral density as given by the Kesten--McKay law (which is correct in the high $T$ limit for the BM model), namely
  \begin{align}
    {\rho^{\rm{RW}} (\lambda) = \frac{2c}{\pi (1-4(\lambda + 1/2)^2)} \sqrt{\frac{c-1}{c^2} - \left(\lambda+ \frac{1}{2} \right)^2}}
\label{kesten}
  \end{align}
Using $c = 3$ and evaluating the principal value integral~\footnote{The principal value integral was evaluated with \texttt{Mathematica}} gives
  \begin{align}
    \hat{I}_2(\lambda) &= \frac{C}{N} \left(1 + \frac{2}{C} \right) +  \frac{C}{N \pi^2 \rho(\lambda)^2}  \left(\frac{1 + 2 \lambda}{4 \lambda (1 + \lambda)} \right)^2
\label{anabolle}
\end{align}
This is our final result. As can be seen, a factor of $\rho(\lambda)^2$ in the denominator remains for the last contribution together with an expression that is symmetric around $\lambda = -1/2$. This clearly accounts for the observations reported in Fig.~\ref{manyTs} (top), where $\hat{I}_2$ is minimal at $\lambda = -1/2$ and grows rapidly when approaching the edges of the spectrum, while the true IPR from DD is flat.
  

 \subsection{Localization of the modes at $T = 0$}
 \label{locsec}

 In Ref.~\cite{tapias2020entropic} it is shown that the symmetrized master operator in the limit $T = 0$ becomes diagonal with elements
 \begin{align}
   M^s_{ij} = -\frac{\delta_{ij} }{c} \sum_{k \neq i} A_{ki} \Theta(E_k - E_i) = -\delta_{ij} \Gamma_i
 \end{align}
  with $\Theta(x)$ the Heaviside step function and $\Gamma_i$ the escape rate from node $i$. This implies that the natural basis $\hat{e}_1 = (1, 0, \cdots, 0),   \hat{e}_2 = (0, 1, 0, \cdots, 0), \ldots$ is the eigenbasis with eigenvalues $-\Gamma_1, -\Gamma_2, \cdots$. In this scenario, direct evaluation of the IPR (eq.~\eqref{empiripr}) yields $1$ for all eigenvectors, which implies full localization.




\end{document}